\documentclass[10pt,aps,twocolumn,prc,superscriptaddress,showpacs,nofootinbib,noshowkeys,floatfix,preprintnumbers]{revtex4-1}
\usepackage{graphicx}
\usepackage[usenames]{color}
\usepackage{amsmath,amssymb}
\usepackage{multirow}
\usepackage{longtable}
\usepackage[normalem]{ulem}
\usepackage{epstopdf}
\usepackage{times}
\usepackage[normalem]{ulem}  

\renewcommand\sout{\bgroup \color{red} \ULdepth=-.5ex \ULset}
\usepackage{bm}
\graphicspath{{./Figs/}}
\begin{document}
\title{Probing $\Omega\Omega$ and $p\Omega$ dibaryons with femtoscopic correlations \\
in relativistic heavy-ion collisions}
\date{\today}
\author{Kenji Morita}
\email{morita.kenji@qst.go.jp}
\affiliation{RIKEN Nishina Center, Wako 351-0198, Japan}
\affiliation{Institute of Theoretical Physics, University of Wroc{\l}aw, PL-50204, Wroc{\l}aw, Poland}
\affiliation{National Institutes for Quantum and Radiological Science and Technology, Rokkasho Fusion Institute, Rokkasho, Aomori, 039-3212, Japan}
\author{Shinya Gongyo}
\affiliation{RIKEN Nishina Center, Wako 351-0198, Japan}
\author{Tetsuo Hatsuda}
\affiliation{RIKEN Interdisciplinary Theoretical and Mathematical Science Program (iTHEMS), Wako 351-0198, Japan}
\affiliation{RIKEN Nishina Center, Wako 351-0198, Japan}
\author{Tetsuo Hyodo}
\affiliation{Yukawa Institute for Theoretical Physics, Kyoto University,
Kyoto 606-8502, Japan}
\affiliation{Department of Physics, Tokyo Metropolitan University,
Hachioji 192-0397, Japan}
\author{Yuki Kamiya}
\affiliation{
CAS Key Laboratory of Theoretical Physics, Institute of Theoretical Physics,
Chinese Academy of Sciences, 
Beijing 100190, China}
\author{Akira Ohnishi}
\email{ohnishi@yukawa.kyoto-u.ac.jp}
\affiliation{Yukawa Institute for Theoretical Physics, Kyoto University,
Kyoto 606-8502, Japan}
\preprint{YITP-19-79, RIKEN-QHP-423, RIKEN-iTHEMS-Report-19}
\begin{abstract}

The momentum correlation functions  of baryon pairs, which reflects
the baryon-baryon interaction at low energies,
are investigated for multi-strangeness pairs ($\Omega\Omega$ and $N\Omega$) 
produced in relativistic heavy-ion collisions.
We calculate the correlation functions based on an expanding source model
constrained by single-particle distributions. 
The  interaction potentials are taken from those obtained from recent lattice
 QCD calculations at nearly physical quark masses.
Experimental measurements of these correlation functions
for different system sizes will help to disentangle the strong interaction
between baryons and to unravel the possible existence of strange dibaryons.
\end{abstract}
\maketitle

\section{Introduction}

Either bound or resonant  dibaryons provide valuable information
on baryon-baryon interactions~\cite{Gal:2015rev,Clement:2016vnl}. 
Historic example is  the  bound deuteron
\cite{urey32:_hydrog_isotop_mass} which indicates  the strong tensor
force in  the $^3SD_1$ nucleon-nucleon interaction
\cite{Rarita:1941zza}. Similarly, observing possible dibaryons with
multi-strangeness would give useful constraints on the unknown
hyperon-nucleon and hyperon-hyperon interactions. The $H$-dibaryon  with
spin $J\!=\!0$ and $S\!=\!-2$~\cite{Jaffe:1976yi}, the $N\Omega$ with
$J\!=\!2$ and $S\!=\!-3$~\cite{Goldman:1987ma,Oka:1988yq}, and the
$\Omega\Omega$ with $J\!=\!0$ and $S\!=\!-6$~\cite{Kopeliovich:1990pp}
are particularly interesting, since the Pauli blocking among valence
quarks do not operate in these systems.

In recent years, ab initio calculations of baryon-baryon
interactions on the basis of lattice quantum chromodynamics (LQCD) 
became possible near the physical quark masses.  This is  due to 
the development of advanced techniques such as
the HAL QCD method~\cite{HALQCD,HALQCD:2012aa}
and the unified contraction algorithm \cite{Doi:2012xd}.
In particular, it was numerically demonstrated that the $\Omega\Omega$
interaction in the $J=0$ channel and the $N\Omega$ interaction in the
$J=2$ channel are attractive enough to hold molecular-like bound states
in the S-wave~\cite{Gongyo:2017fjb,Iritani:2018sra}. 

To study such  multi-strangeness systems experimentally, high-energy
heavy-ion collisions provide a unique opportunity allowing direct search
via invariant mass spectrum \cite{ExHIC,Cho:2017dcy} as
well as indirect search via momentum correlations
\cite{Morita:2014kza,Morita:2016auo,Ohnishi:2016elb,Hatsuda:2017uxk,Cho:2017dcy}. 
As for the latter, a ratio of the correlation functions obtained from
different source sizes  has been  theoretically introduced and called
``small-to-large (SL) ratio''~\cite{Morita:2016auo}. 
This is useful to access  e.g.  the strong $p\Omega$ interaction  without
much contamination from  the Coulomb interaction  at small relative
momentum.  Subsequently,  the measurement of the momentum correlation of
$p\Omega$  was conducted in Au+Au collisions at RHIC
\cite{STAR:2018uho}.

The main purpose of this paper is to study the pair momentum correlation
functions of the dibaryon candidates,
$\Omega\Omega$ and  $p\Omega$,  by extending our previous analysis
~\cite{Morita:2016auo,Morita:2014kza,Ohnishi:2016elb,Hatsuda:2017uxk,Cho:2017dcy}.
We employ the latest interactions obtained from the (2+1)-flavor
lattice QCD simulations with nearly physical quark masses~\cite{Gongyo:2017fjb,Iritani:2018sra}.
Also we use an expanding source model constrained by experimental
transverse momentum spectra and multiplicities.
In  Sec.~\ref{sec:HBT}, we recapitulate the general feature
of the momentum correlation function in a simplified example to give an
account of how the final state interaction (FSI)
is translated into the pair correlations. 
A model for the emission source function is described in
Sec.~\ref{sec:source}. We give details of the potential and resultant
correlation functions for $\Omega\Omega$ pairs and $p\Omega$ pairs
in Sec.~\ref{sec:WW} and \ref{sec:NOmega}, respectively.
Section \ref{sec:conclusion} is devoted to summary and concluding remarks.
In Appendix A, the system size dependence of the momentum correlation 
for $p\Omega$ with uncertainty quantification are examined.
In Appendix B, we show a comparison of 
the $p\Omega$ potential in \cite{Morita:2016auo}
 with that in ~\cite{Iritani:2018sra} adopted in the present paper.

\section{Two-particle momentum correlation from final state
 interactions}
 \label{sec:HBT}

\subsection{Formalism}
We briefly recapitulate the general property of the two-particle
momentum correlation function with FSI.
More details can be found in, e.g., Refs.~\cite{Cho:2017dcy,Lisa:2005dd}.

The momentum correlation function between particles 1 and 2 with
respective momenta $p_1$ and $p_2$ is defined by the
ratio of two-particle spectrum 
$N_{12}(\bm{p}_1,\bm{p}_2)=E_1E_2 dN_{12}/d\bm{p}_1d\bm{p}_2$ 
and the  product of single-particle spectra $N_i(\bm{p}_i)=E_{i} dN_i/d\bm{p}_i$ as 
\begin{equation}
C(q^\mu,P^\mu)
= \frac{N_{12}(\bm{p}_1,\bm{p}_2)}{N_1(\bm{p}_1)N_2(\bm{p}_2)},\label{eq:c2_general}
\end{equation}
with $E_i = \sqrt{\bm{p}_i^2+m_i^2}$
being the on-shell particle energy.
The center-of-mass momentum $P$ and the generalized relative momentum $q$
are defined by
\begin{align}
 P^\mu &= p_1^\mu+p_2^\mu, \\
 q^\mu &= \frac{1}{2} \left[ p_1^\mu-p_2^\mu - \frac{(p_1-p_2)\cdot
 P}{P^2}P^\mu \right].
\end{align}
One may, in principle, measure the correlation function as a function of
three independent components of the relative momentum $q^\mu$.
Such a decomposition has been utilized to investigate expansion dynamics of the
hot matter through pion correlations \cite{Lisa:2005dd}. 
In practice, particles except for  pions do not allow for such detailed
study due to limited statistics. Hereafter, we consider only
one-dimensional correlation function with respect to the invariant
relative momentum $q =\sqrt{-q_\mu q^\mu}$.
Then we can define the experimental correlation function by 
\begin{equation}
 C(q) = \frac{A_{12}(q)}{B_{12}(q)}\label{eq:pair},
\end{equation}
where $A_{12}(q)$ is for the number of pairs from the same event
while $B_{12}(q)$ is constructed from mixed events.
Eq.~\eqref{eq:pair} is related to the two-particle and single-particle
spectra as
\begin{equation}
 C(q) = \frac{\displaystyle \int \frac{d\bm{p}_1}{E_1}
  \frac{d\bm{p}_2}{E_2} N_{12}(\bm{p}_1,\bm{p}_2)\delta(q-\sqrt{-q^2})}{\displaystyle \int \frac{d\bm{p}_1}{E_1}
  \frac{d\bm{p}_2}{E_2}N_1(\bm{p}_1)N_2(\bm{p}_2)\delta(q-\sqrt{-q^2})}\label{eq:c2_projected} ,
\end{equation}
where the momentum integration should reflect 
the experimental momentum coverage.

The source function $S_i(x,\bm{p})$ is defined as the phase space distribution 
of the particles at freeze-out and is related to the single-particle spectrum as
\begin{equation}
 N_i(\bm{p}) = \int d^4 x S_i(x,\bm{p}).\label{eq:single-source}
\end{equation}
Then  the two-particle spectrum from uncorrelated (chaotic) sources reads
\begin{align}
 N_{12}&(\bm{p}_1,\bm{p}_2) \nonumber\\
 &\simeq \int d^4 x d^4 y
 S_1(x,\bm{p}_1)S_2(y,\bm{p}_2)|\Psi(x,y,\bm{p}_1,\bm{p}_2)|^2
\label{eq:chaotic}\\
 &\simeq \int d^4x d^4 y S_1(x,\bm{p}_1)S_2(y,\bm{p}_2)|\varphi(\bm{q}^*,\bm{r}^*)|^2,\label{eq:2particledistribution}
\end{align}
where $\Psi(x,y,\bm{p}_1,\bm{p}_2)$ denotes
the Bethe-Salpeter amplitude describing propagations of pairs from the
emission point $x$ and $y$ to the asymptotic state with momenta
$\bm{p}_1$ and $\bm{p}_2$. The squared two-particle amplitude is
well approximated by the relative wave function
$\varphi(\bm{q}^*,\bm{r}^*)$ in the pair rest frame defined by $\bm{P}=0$.
Here $\bm{q}^*$ and
$\bm{r}^*=\bm{x}^*-\bm{y}^*$ are the spatial
components of relative momentum and the relative coordinate defined in
the pair rest frame, respectively.
 Note that $q=|\bm{q^*}|$ when $\bm{P}=0$.
The information on the pairwise interaction is encoded in
$\varphi(\bm{q}^*,\bm{r}^*)$ 
which can be obtained by solving the Schr\"{o}dinger equation. 
The squared relative wave function $|\varphi(\bm{q}^*,\bm{r}^*)|^2$
can be viewed as a weight factor for the two-particle emission. 
Therefore, $N_{12}(\bm{p}_1,\bm{p}_2)$ reduces to the product
$N_1(\bm{p}_1)N_2(\bm{p}_2)$ for $|\varphi(\bm{q}^*,\bm{r}^*)|^2 = 1$.
Note that
Eq.~\eqref{eq:chaotic} is valid under the chaotic source assumption,
the so-called smoothness assumption
($S_i(x,\bm{p})$ being smooth in the momentum space),
and the negligible correlation with other particles.
The validity of
Eq.~\eqref{eq:2particledistribution}
 further requires $\bm{q}^*$ to be small 
 compared with the particle masses in order for $\varphi(\bm{q}^*,\bm{r}^*)$
to be regarded as the relative wave function.
(See Refs.~\cite{Femto} for detailed discussion.)

If the center-of-mass coordinate and relative time are integrated,
we obtain the Koonin-Pratt formula,
\begin{align}
 C^{(\rm KP)}(q) = \int d\bm{r}^*
  S_{12}^\text{rel}(\bm{r}^*)|\varphi(\bm{q}^*,\bm{r}^*)|^2\label{eq:KP}
  ,
\end{align}
where the relative source function
$S_{12}^\text{rel}(\bm{r}^*)$ can be viewed as the relative source
distribution in the pair rest frame.
The relative source function is momentum dependent when the emission
point is correlated with momentum, as is the case for collective
expansion. 

In this paper, we adopt a parameterized model of $S_i(x,\bm{p})$
with hydrodynamic expansion~\cite{expansion}
with the parameters constrained from single-particle spectra through
Eq.~\eqref{eq:single-source}.
Detailed analyses of $\pi$-$\pi$ correlations at RHIC have revealed that
various features of the expanding matter need to be implemented to
produce the pion emitting source compatible with measurements
\cite{Pratt:2008qv}. Therefore, our parameterized source may be an oversimplification. 
On the other hand, precise shape of the source function is not crucially
important in our  one-dimensional correlation.  Use of more realistic
source functions through the implementations of state-of-the-art
dynamical models will be left for future studies.

\subsection{Correlations from S-wave scattering}

\begin{figure}[!t]
 \centering
 \includegraphics[width=\columnwidth]{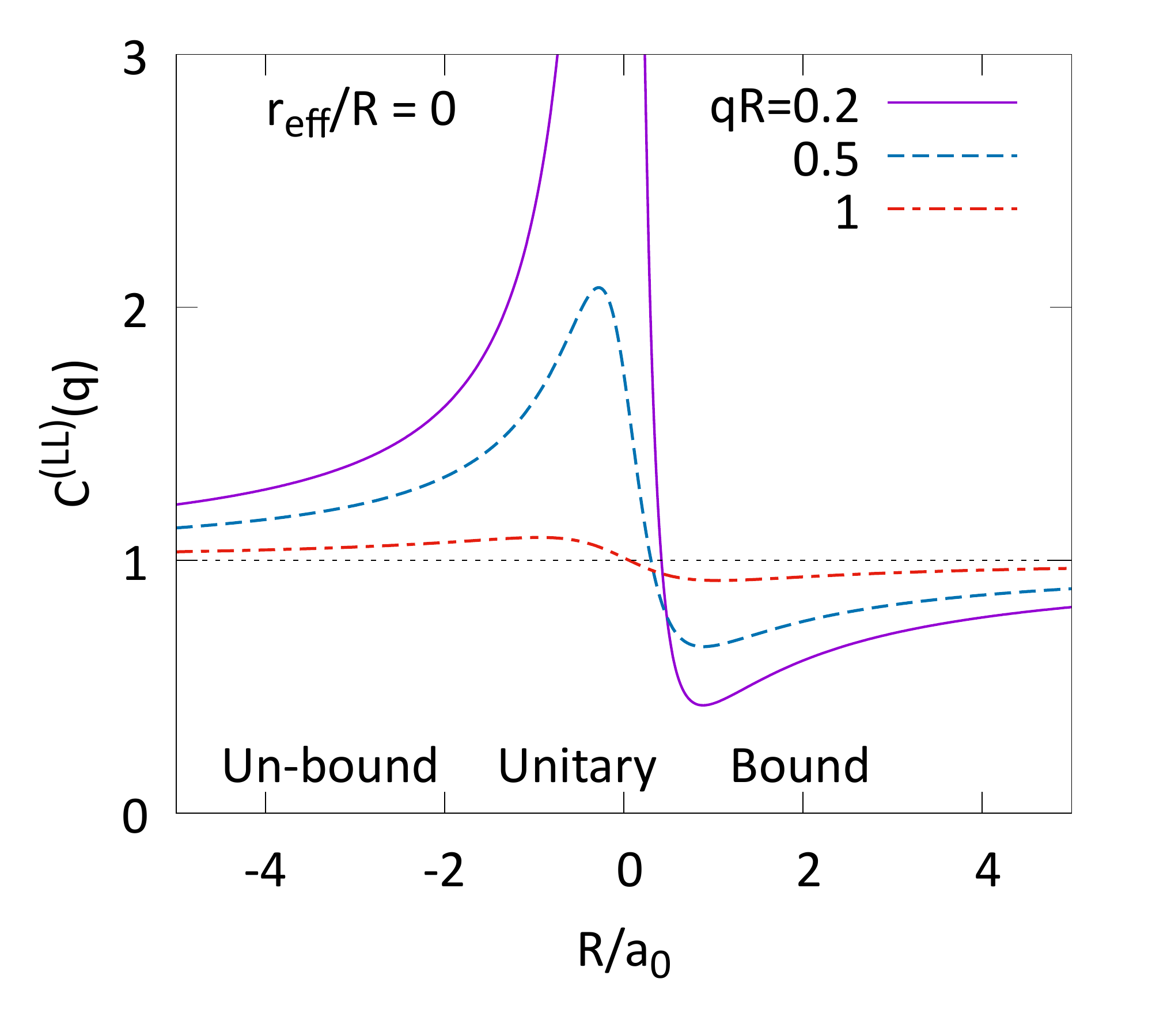}\\
 \includegraphics[width=\columnwidth]{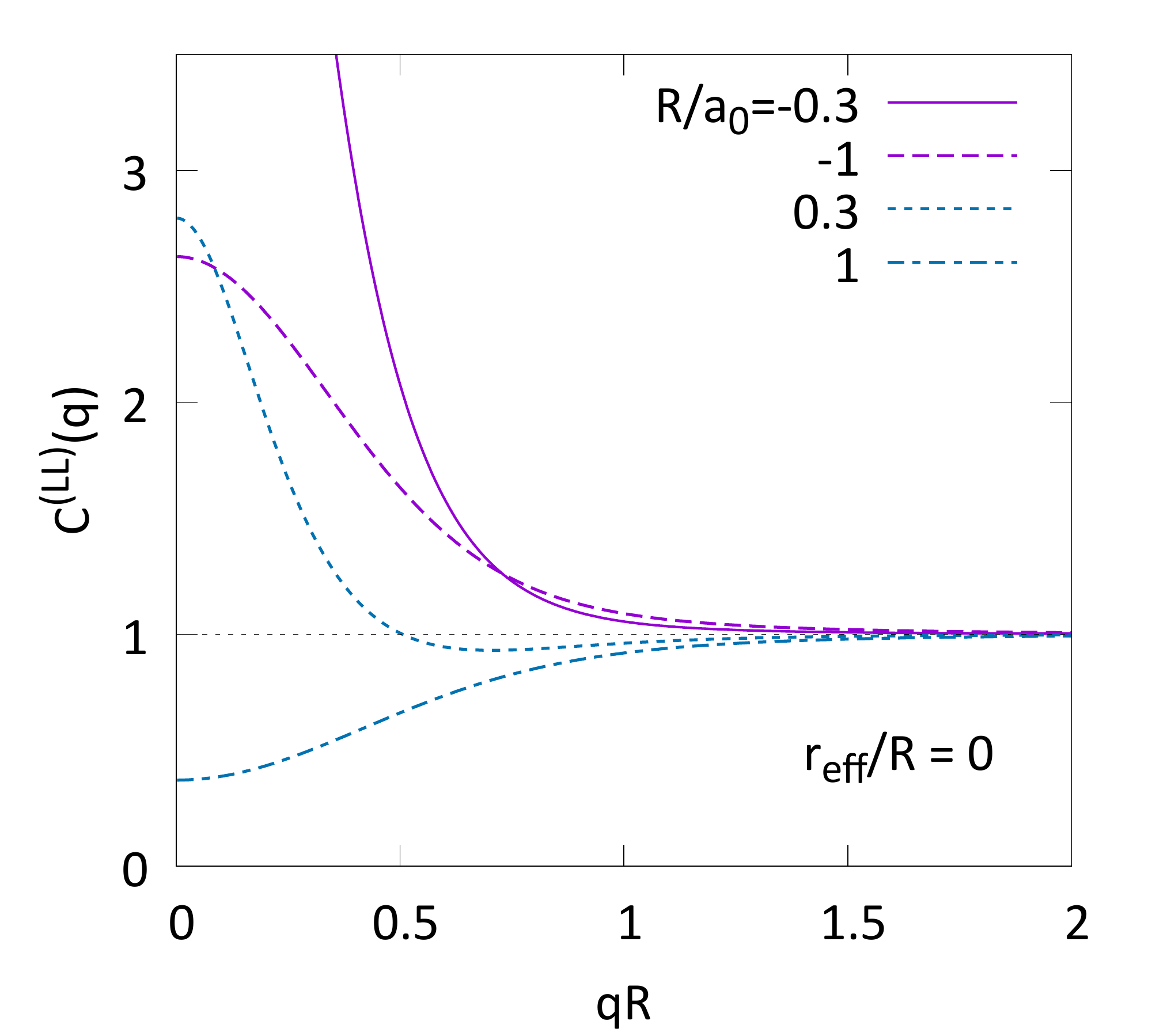}
 \caption{The correlation function  $C^{(\text{LL})}(q)$ with $r_{\rm
 eff}=0$  as a function of $R/a_0$ for different $qR$ (upper panel) and
 as a function of $qR$ for different $R/a_0$ value (lower
 panel). In the present sign convention,  $a_0 > 0$ corresponds to
 the existence of a bound state. }
 \label{fig:LL}
\end{figure}

Owing to the short-range nature of the strong interaction, the modification
of the relative wave function of non-identical particle pairs
takes place mainly in the S-wave state.
Thus, one may express
\begin{equation}
 \varphi(\bm{q},\bm{r}) = e^{i\bm{q}\cdot\bm{r}} - j_0(|\bm{q}| r) + \psi_{|\bm{q}|}(r)\label{eq:wf_int}
 ,
\end{equation}
where
$j_0(x)$ is the zeroth-order spherical Bessel function,
and $\psi_{|\bm{q}|}(r)$ is the S-wave relative wave function
with the pairwise interaction effects.
The connection of the pairwise interaction with the correlation function
can be nicely illustrated by employing a static and spherically
symmetric source function, 
$S_{12}^{\text{rel}}(\bm{r^*}) = S(r=|\bm{r^*}|)$, as ~\cite{Morita:2016auo}
\begin{align}
C^{\rm (KP)} (q) = 1+ \int [d\bm{r^*}]\left( 
 |\psi_{q}(r)|^2 - |j_0(qr)|^2
\right) \label{eq:corr_static}
,
\end{align}
where $[d\bm{r^*}] = d\bm{r^*} S(r)$ with $S(r)$ being properly
normalized as 
$\int [d\bm{r^*}] = 1$. One immediately finds that the deviation
of the wave function from the non-interacting one is directly
translated into the correlation function and that the relative source
function acts as a weight factor
at relative distance $r$. 

Furthermore, when the source size is not too small compared to the
interaction range, the integral is dominated by the contribution outside
the interaction range such that the wave function can be approximated by
its asymptotic
form $\psi_q(r) \sim e^{-i\delta} \sin(qr + \delta)/(qr)$ with $\delta$
being the S-wave scattering phase shift. Employing a Gaussian source 
$S(r)\propto \exp(-r^2/4R^2)$ and the effective range formula for small $q$,
\begin{equation}
 q\cot\delta \simeq -\frac{1}{a_0} + \frac{1}{2}r_{\text{eff}}q^2,
\end{equation}
one can express the correlation function 
in terms of the scattering
length $a_0$ and the effective range $r_{\text{eff}}$, which is known as 
the Lednick\'{y}-Lyuboshits (LL) formula~\cite{lednicky82:_influence},
\begin{align}
 C^{(\text{LL})}(q) &= 1 + \frac{|f(q)|^2}{2R^2}
 F_3\left( \frac{r_{\text{eff}}}{R}
 \right) + \frac{2\text{Re}f(q)}{\sqrt{\pi}R}F_1(2qR) \nonumber\\
  & \quad -\frac{\text{Im}f(q)}{R}F_2(2qR).
\end{align}
Here  $f(q) = (q\cot\delta-iq)^{-1}$ is the scattering amplitude, 
$F_1(x)=\int_{0}^{x}dt e^{t^2-x^2}$, $F_2(x)=(1-e^{-x^2})/x$, and
$F_3(x) = 1-x/(2\sqrt{\pi})$.
Since the scattering length dominates the behavior of the phase shift at
small $q$, this correlation function is mainly determined by the
scattering length and the source size: For $r_{\text{eff}}=0$, 
$C^{(\text{LL})}(q)$ is a function of two dimensionless variables,
$qR$ and $R/a_0$~\cite{Cho:2017dcy}.

Figure \ref{fig:LL} represents characteristics of
the correlation function  $C^{(\text{LL})}(q)$ with $r_{\rm eff}=0$.
For a fixed $qR$ (upper panel),
the correlation function exhibits
non-monotonic changes against the ratio of the system size to the
scattering length. It shows a strong peak around $R/a_0 \sim 0$ for small $qR$
due to the strong enhancement of the wave function.
We call the region where $C(q)$ is enhanced as the ``unitary region" 
throughout this paper.
The peak is smeared as $qR$ is increased.
As the attraction becomes weaker ($a_0 < 0$), the
correlation is also weakened to exhibit monotonic decrease
with decreasing $R/a_0$ and increasing $qR$.
On the other hand, if the attraction is strong enough to
accommodate a bound state ($a_0 > 0$), $C(q)$ rapidly decreases with $R/a_0$
then takes values less than unity implying the depletion of correlated pairs
at small $qR$. 
The depletion can be understood by so-called the structural core;
the scattering wave function needs to be orthogonal to the bound state
wave function, then it has a node in the interaction range
as if there is a repulsive core.
Thus the squared wave function is suppressed on average.

The above properties of $C(q)$ are essential in order to extract the pairwise
interaction from the measured correlation functions.
In particular, the behavior of $C(q)$ for different system size provides
detailed information on the scattering parameters
as shown in the lower panel of Fig.~\ref{fig:LL}.
Consider the case where  $C(q) \gg 1$ at small $qR$.
It indicates that the system is in the unitary region where $|R/a_0|$ is small,
while the sign of $a_0$  is unknown.
However, by increasing $R$ with $a_0$ and $qR$ fixed,
$C(q)$ eventually becomes smaller than 1 for positive $a_0$,
while $C(q)$ is always larger than 1 for negative $a_0$.

In reality, the correlation at small $q$ originates not only from the
single-channel FSI but also from the quantum statistics in the
case of identical pairs (HBT effect),
from the Coulomb interaction, and from the coupled channel effect
\cite{Haidenbauer:2018jvl}.  Furthermore, the correlation from the
HBT effect is affected by the collective flow through the modification
of the source geometry. As a result, 
even for non-identical pairs, the absolute  magnitude of $C(q)$ with
respect to unity is not always a useful measure to quantify the effect
of FSI in heavy-ion collisions. However, by taking a ratio of the
correlation functions with small and large system sizes as
\begin{equation}
 C_\text{SL} (q) =  C_\text{small-R} (q)/C_\text{large-R} (q),
\end{equation}
one can nicely cancel out the effect of the Coulomb
interaction between charged pairs and  extract the  FSI from the strong interaction,
as demonstrated in \cite{Morita:2016auo}. We will follow this idea in this paper
 to study  $\Omega\Omega$ and $p\Omega$ correlations.

\section{Modeling emission function}\label{sec:source}

\begin{figure}[!t]
 \includegraphics[width=\columnwidth]{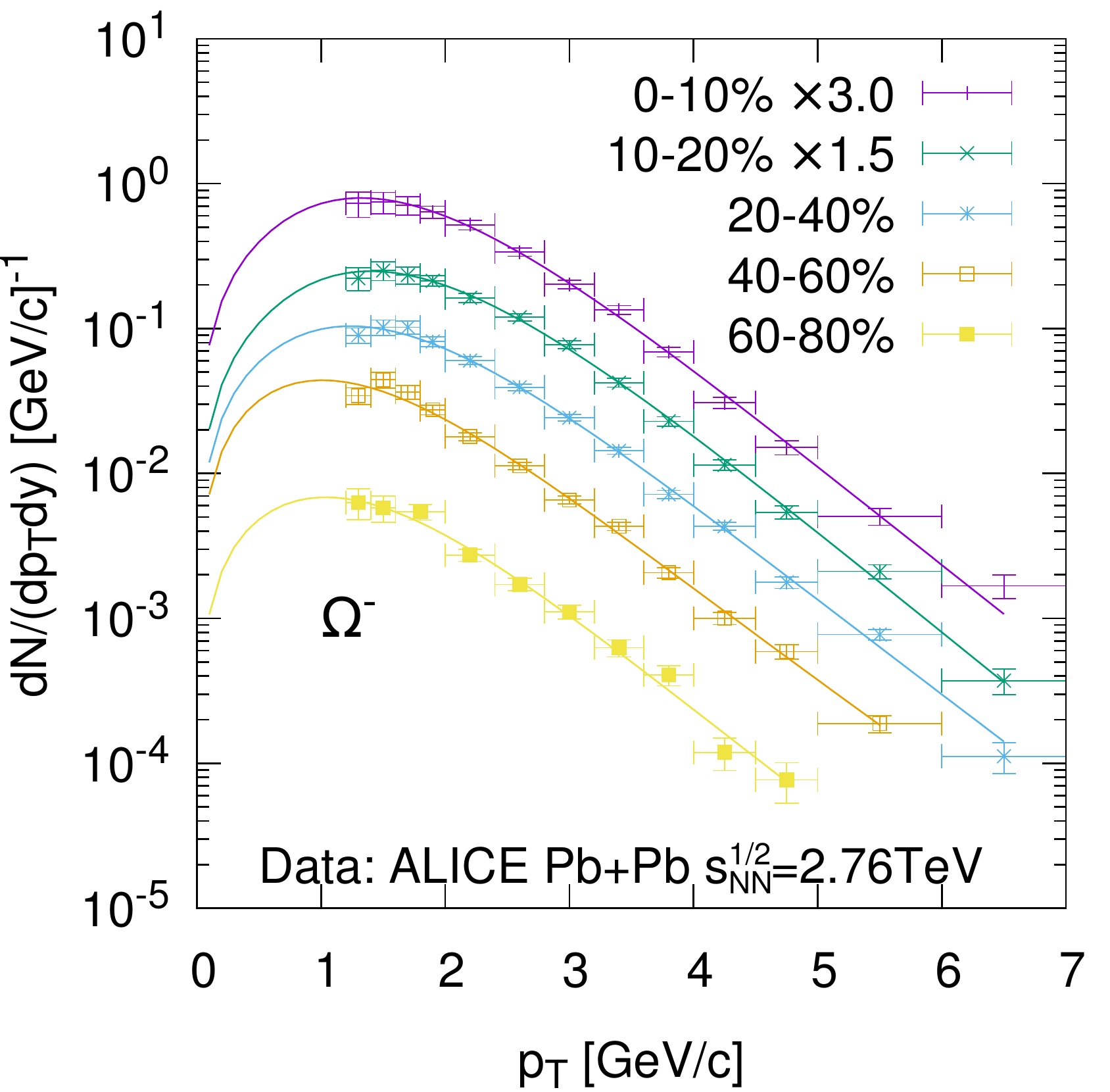}
 \includegraphics[width=\columnwidth]{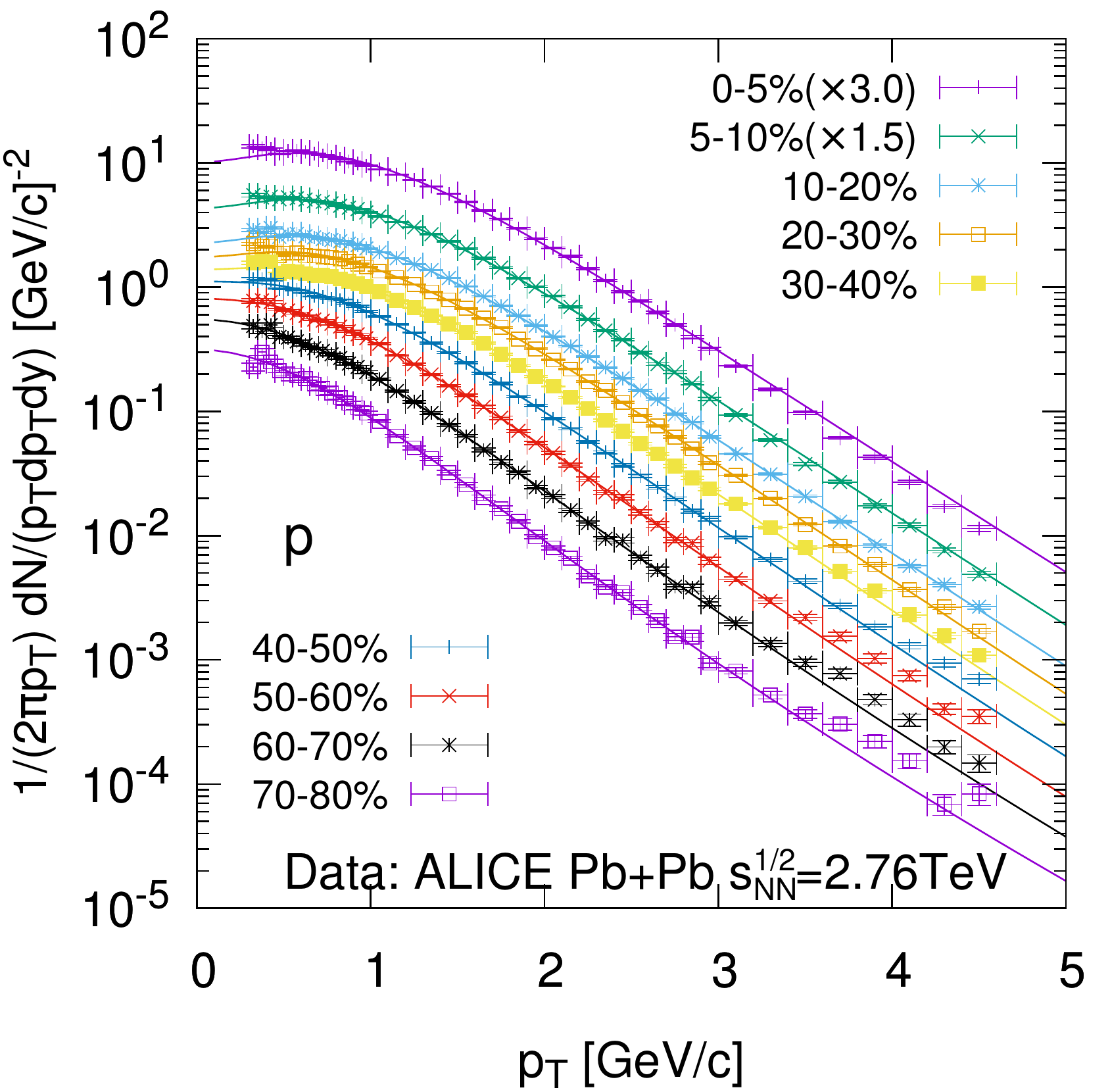}
 \caption{Transverse momentum spectra of 
$\Omega$ (upper)
   and $p$ (lower). Experimental data are taken from
 \cite{ABELEV:2013zaa} and \cite{Abelev:2013vea} for $\Omega$ 
 and protons, respectively. Two most central events are scaled by factor
 3 and 1.5 for better comparison.}
 \label{fig:ptspectra}
\end{figure}

As seen from Fig.~\ref{fig:LL}, the correlation from FSI strongly
depends on the source size. In order to extract the pairwise
interaction from the correlation function,
one needs to know the source size or to look
at the system size dependence of the correlation \cite{Morita:2016auo}.
Therefore, modeling the particle source is one of the indispensable
ingredients in quantitative analyses. Here, we employ a thermal source
model with hydrodynamic expansion in which parameters are so tuned as to
reproduce relevant particle yields and spectra. 

We assume that the baryon production takes place at chemical and thermal
freeze-out temperature $T_f$ from a cylindrically expanding
boost-invariant fireball, where the flow velocity $u^\mu(x)$ is parameterized as
$u^\mu = (\cosh\eta_s \cosh y_T, \sinh y_T \cos\phi, \sinh y_T \sin\phi,\sinh\eta_s \cosh y_T)$ 
with $\eta_s = \tanh^{-1}(z/t)$ being the spacetime rapidity. 
The transverse rapidity $y_T$ is parameterized as 
$y_T = \alpha (r_T /R_T)^\beta$, where $\alpha$ are $\beta$ are the fitting
parameters and $R_T$ denotes the transverse source size.
Then the emission function of particle species $i$
can be written as~\cite{expansion}
\begin{equation}
 d^4x S_i(x,\bm{p}) = \tau_0 d\eta_s d^2r_T
  \frac{d}{(2\pi)^3} n_\text{F}(u\cdot p,T)
  \exp\left(-\frac{r_T^2}{2R_T^2}\right)
  ,
  \label{eq:expandingsource}
\end{equation}
where
$\bm{p}$ is the on-shell momentum, $x$ is the spacetime emission point,
$d$ denotes the spin degeneracy,
and $n_\text{F}$ denotes the Fermi distribution function. 
 We assume that hadrons are produced
at a constant proper time $\tau=\sqrt{t^2-z^2}=\tau_0$ with 
a Gaussian profile in the transverse direction.
The use of azimuthally symmetric profile is an oversimplification since it does
not account for the significant anisotropic flow in non-central events,
but we retain it in order to reduce the number of parameters. In fact, 
the one-dimensional baryon-baryon correlation functions are not expected
to be strongly sensitive to detailed source shape in the transverse
plane, since it can be expressed in terms of relative source
distribution \eqref{eq:KP}. By integrating over $\eta_s$ and $r_T$,
one obtains the single particle spectrum, $EdN/d^3\bm{p}$. In the
Boltzmann approximation $m \gg T$, the thermal spectrum is proportional
to the volume factor $V=2\pi\tau_0 R_T^2$, so that we have 
\begin{align}
 \frac{dN}{dy p_T dp_T 2\pi} &= \frac{d}{(2\pi)^3}2m_T V
  \int_0^\infty d\rho e^{-\rho^2/2} \nonumber \\
 & \quad \times I_0\left(\frac{p_T}{T}\sinh y_T\right)K_1
   \left(\frac{m_T}{T}\cosh y_T \right) ,
\end{align}
where $I_0$ and $K_1$ are the modified Bessel functions.

The parameters in our model are determined by the following procedure.
First, we fix the freeze-out temperature to $T_f=155$ MeV from the
fit to the various particle multiplicity data  at LHC
\cite{Andronic:2017pug}. 
We perform a fit
to the experimental transverse momentum spectra of each species
 by varying three parameters ($V, \alpha$ and $\beta$).
Finally we fix $\tau_0=10$ fm/$c$ from a
freeze-out temperature in a hydrodynamic model calculation
\cite{Zhu:2015dfa} for the most central event bin (5-10\%) in 
 $\Omega$
production analyses \cite{ABELEV:2013zaa}. We take the relation
$\tau_0 \simeq (dN/dy)^{1/3}$ which is expected from the property of longitudinal
HBT radii $R_{\text{long}}\simeq \tau_0 \sqrt{T_f/m_T}$
\cite{Makhlin:1987gm} and well-established relation between the HBT
radii and multiplicity. Then $R$ is obtained from the fitted values of
the volume factor $V$.

Fig.\ref{fig:ptspectra} displays the fitted transverse momentum spectra
for  $\Omega$s and protons.
The obtained parameter sets are summarized in Table \ref{tbl:source_parameters}.
We take into account two-body decay contributions from resonances with mass $m_R < 2$ GeV 
to the proton spectra. 
We note that those resonance contributions are important to fit the total yield
of protons with reasonable system sizes. 
Note also that there is so-called thermal proton
yield anomaly at LHC \cite{Andronic:2017pug}. (See
Ref.~\cite{Andronic:2018qqt} for a possible resolution.) 
The proton spectra have more detailed centrality bins than those of the
 $\Omega$, such that fits are made for those data.
In the calculations of the
correlation function below, we adjust the centrality selections to
$\Omega$ data. Thus, the parameters shown in Table
\ref{tbl:source_parameters} are those used in the subsequent
calculations and are obtained by averaging over corresponding
centralities in the spectrum. (i.e., 0-10\% parameters are obtained by
averaging 0-5\% and 5-10\% with multiplicity being the weight.)
Clearly, the present model is too simple to fully account for other
possible contributions to the proton spectrum such as rescattering
effect after chemical freeze-out. Nevertheless, we have checked that
proton HBT radii from the model are consistent with measurements
\cite{Adam:2015vja}. Therefore, we expect the following results remain
valid for more realistic modeling of the particle sources.

\begin{table}[h]
 \caption{Parameters in the emission function \eqref{eq:expandingsource}
 for different centralities and particle species.}
 \label{tbl:source_parameters}
 \begin{tabular}{c|cccccccc}\hline
   Centrality & $\tau_0$ [fm/$c$] & $R_T^{\Omega}$ [fm] &
  $R_T^{p}$ & $\alpha^\Omega$ & $\beta^\Omega$ &  $\alpha^p$ & $\beta^p$ \\\hline
   $0-10\%$  & 10.0  & 8.0  & 6.8  & 0.584 & 0.628 & 0.759 & 0.421 \\ 
   $10-20\%$ & 9.085 & 6.75 & 6.23 & 0.618 & 0.579 & 0.750 & 0.425 \\
   $20-40\%$ & 7.5   & 5.88 & 5.2  & 0.546 & 0.692 & 0.707 & 0.466 \\
   $40-60\%$ & 5.5   & 4.38 & 3.92 & 0.444 & 0.858 & 0.604 & 0.6 \\
   $60-80\%$ & 3.62  & 2.12 & 2.66 & 0.456 & 0.812 & 0.456 & 0.82 \\ \hline
 \end{tabular}
\end{table}

\section{$\Omega\Omega$ correlation}
\label{sec:WW}

First we discuss
pairs of $\Omega(1672)$ particles. A recent LQCD calculation shows
that the $J=0$ $\Omega\Omega$ system has a shallow bound state
\cite{Gongyo:2017fjb}.  Direct detection of the $\Omega\Omega$
dibaryons (di-Omega) is highly challenging because of the tiny production rate
for the $S=-6$ object even in heavy-ion collisions and the background
yields of the decay products would be high.  On the other hand,
the high luminosity upgrade
at the LHC may allow for measuring the momentum correlation of
$\Omega\Omega$ pairs in the future.  

\subsection{$\Omega\Omega$ interaction from lattice QCD}
\label{subsec:OO-L}

Since $\Omega$ has a spin $3/2$, the $\Omega\Omega$ pairs can have
$J=0,1,2$ and 3.
Among others, the $J=0$ state  is expected to have appreciable S-wave
attraction without suffering from the Pauli exclusion effect for valence
quarks. The interaction potential $V_{\Omega\Omega}^{J=0}$  was recently
calculated by (2+1)-flavor lattice QCD simulations \cite{Gongyo:2017fjb}
with a large lattice volume (8.1 fm)$^3$, a small lattice spacing 
$a \simeq$ 0.0846 fm and nearly physical quark masses 
($m_{\pi} \simeq$ 146 MeV, $m_{K}\simeq$525 MeV, $m_{N}\simeq$964 MeV,
and $m_{\Omega}\simeq$1712 MeV).
In the time-dependent HAL QCD method ~\cite{HALQCD:2012aa} employed in
the analysis, the lattice data at moderate values of the Euclidean
time, $t \sim (1-2)~\mathrm{fm}$ are found to be sufficient to extract
the baryon-baryon interaction. For $\Omega\Omega$, the  interval
$t/a=16-18$ is chosen to avoid the contamination from  the excited state
of a single  $\Omega$  at small $t$ and  large statistical errors at
large $t$.

\begin{figure}[!t]
 \centering
 \includegraphics[width=\columnwidth]{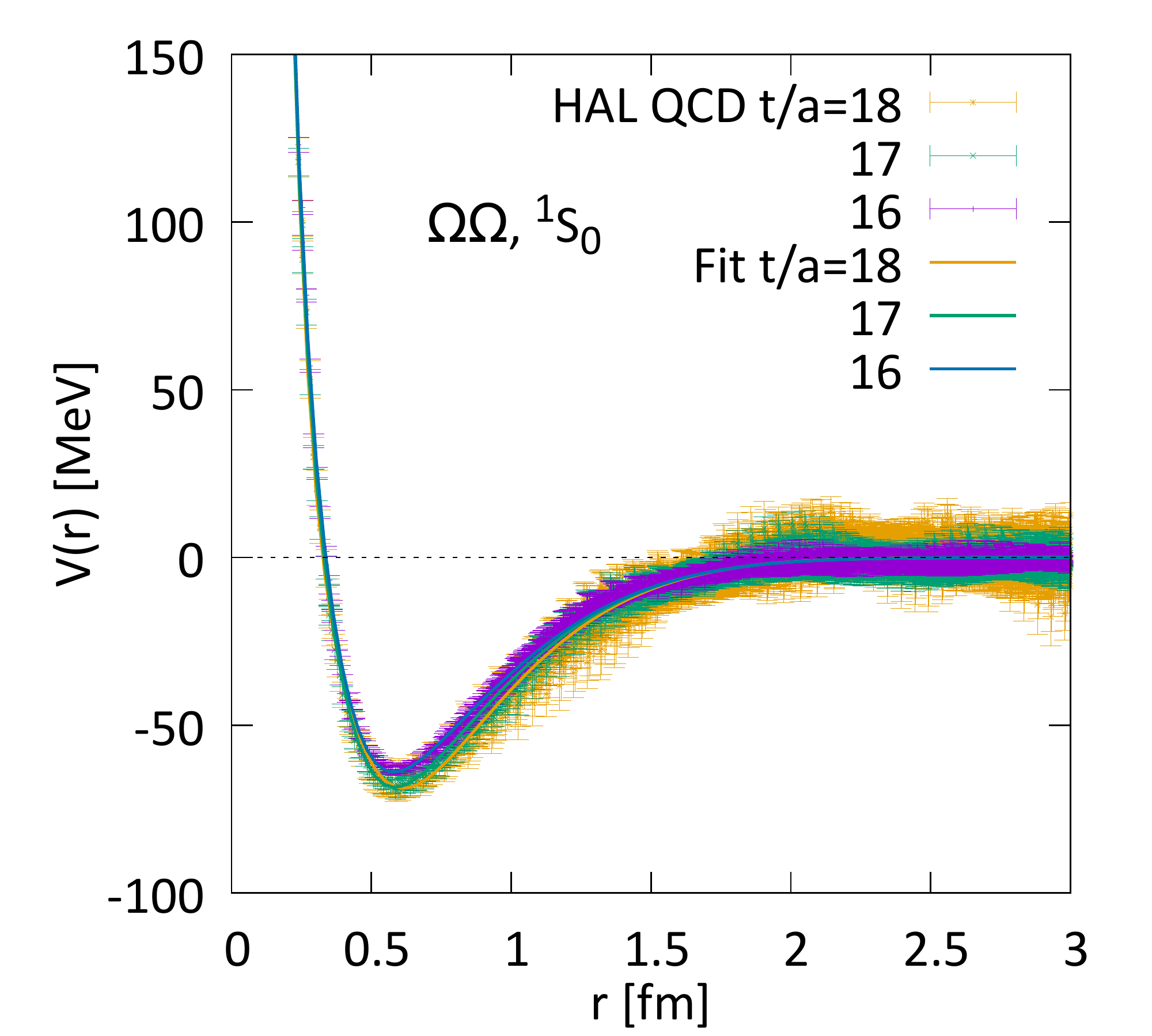}
 \caption{The $\Omega\Omega$ potential in $J=0$ channel from lattice QCD simulations
 \cite{Gongyo:2017fjb}. The lattice data are fitted by the form, 
 $V_\text{fit} (r)= \sum_{j=1,2,3} c_j e^{-(r/d_j)^2}$.}
 \label{fig:pot_omega}
\end{figure}

Resultant potentials with statistical errors are recapitulated in
Fig.~\ref{fig:pot_omega} together with the fitted potential of the
3-range Gaussian form ~\cite{Gongyo:2017fjb}. The scattering length and
the effective range without the Coulomb repulsion are $a_0 \simeq $ 4.6
fm and $r_\text{eff} \simeq$ 1.27 fm, respectively,  so that a weakly
bound di-Omega appears with the binding energy  $E_B \simeq $ 1.6 MeV.

Table \ref{tbl:WW_scattering} shows the low energy scattering
parameters and binding energies obtained by solving
the Schr\"{o}dinger equation in the presence of the attraction from the
strong interaction and the repulsion from the Coulomb interaction.
The already large positive scattering
length found in lattice QCD calculations is
further driven toward the unitary limit ($a_0 \gg r_{\rm eff}$) by the Coulomb repulsion.
The obtained scattering length exceeds the effective source size in heavy-ion
collisions, therefore one can expect the correlation function belongs to the unitary region 
characterized by $R/a_0 \sim 0$ in Fig.~\ref{fig:LL}. 

\begin{table}[h]
 \caption{Scattering length $a_0$, effective range $r_{\text{eff}}$, and
 binding energy of the $\Omega\Omega$ pair with the lattice QCD potential for
 different $t/a$ and the Coulomb repulsion.}
 \label{tbl:WW_scattering}
 \begin{tabular}[t]{c|ccc}\hline
  $t/a$ & $a_0$ [fm] & $r_{\text{eff}}$ [fm] & $E_B$ [MeV]\\\hline
  16    & 65.28 & 1.29 & 0.1\\
  17    & 17.59 & 1.24 & 0.54 \\
  18    & 11.69 & 1.26 & 1.0 \\ \hline
 \end{tabular}
\end{table}

\subsection{Correlation function}

\begin{figure*}[t]
 \includegraphics[width=0.32\textwidth]{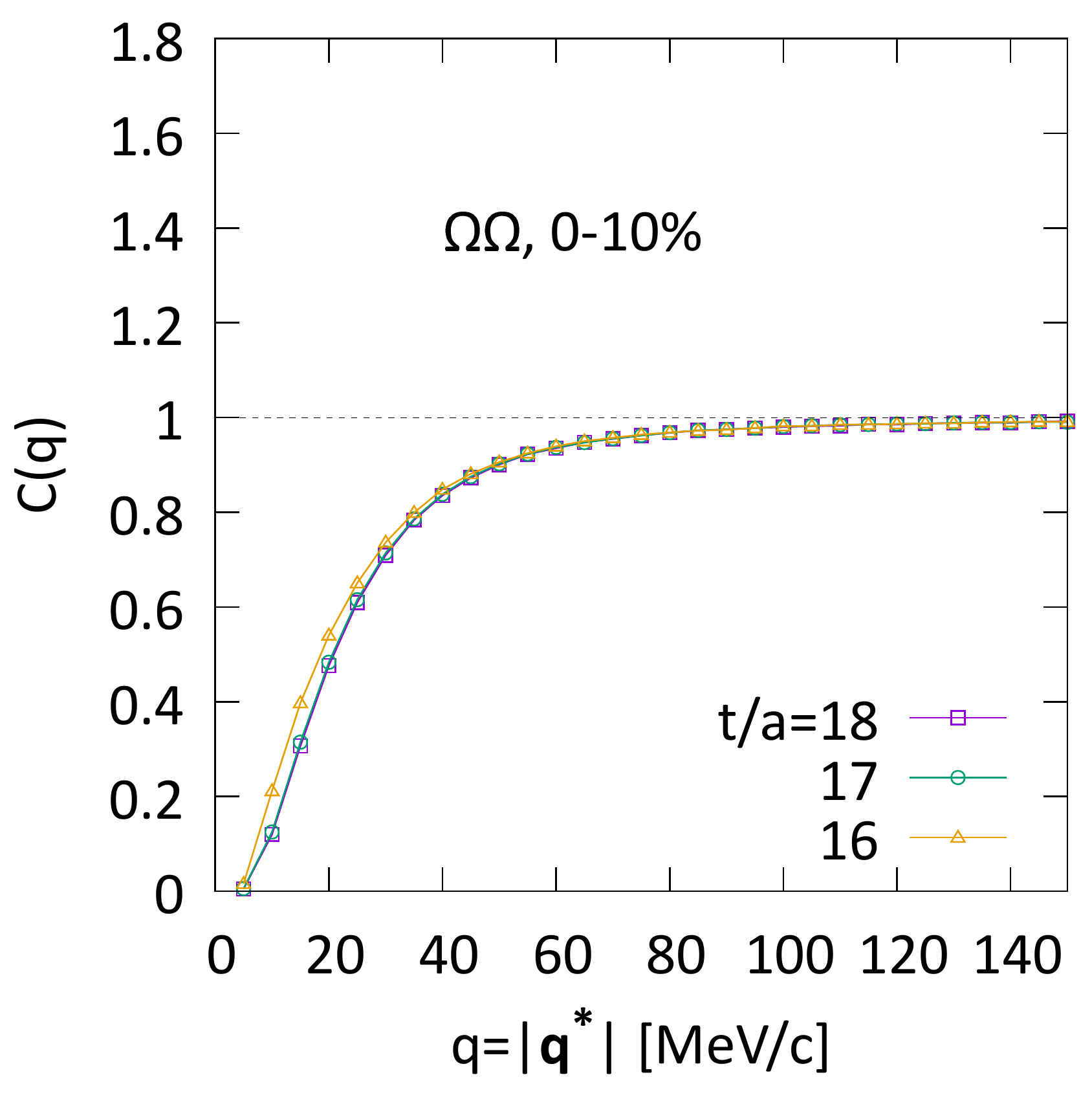}
 \includegraphics[width=0.32\textwidth]{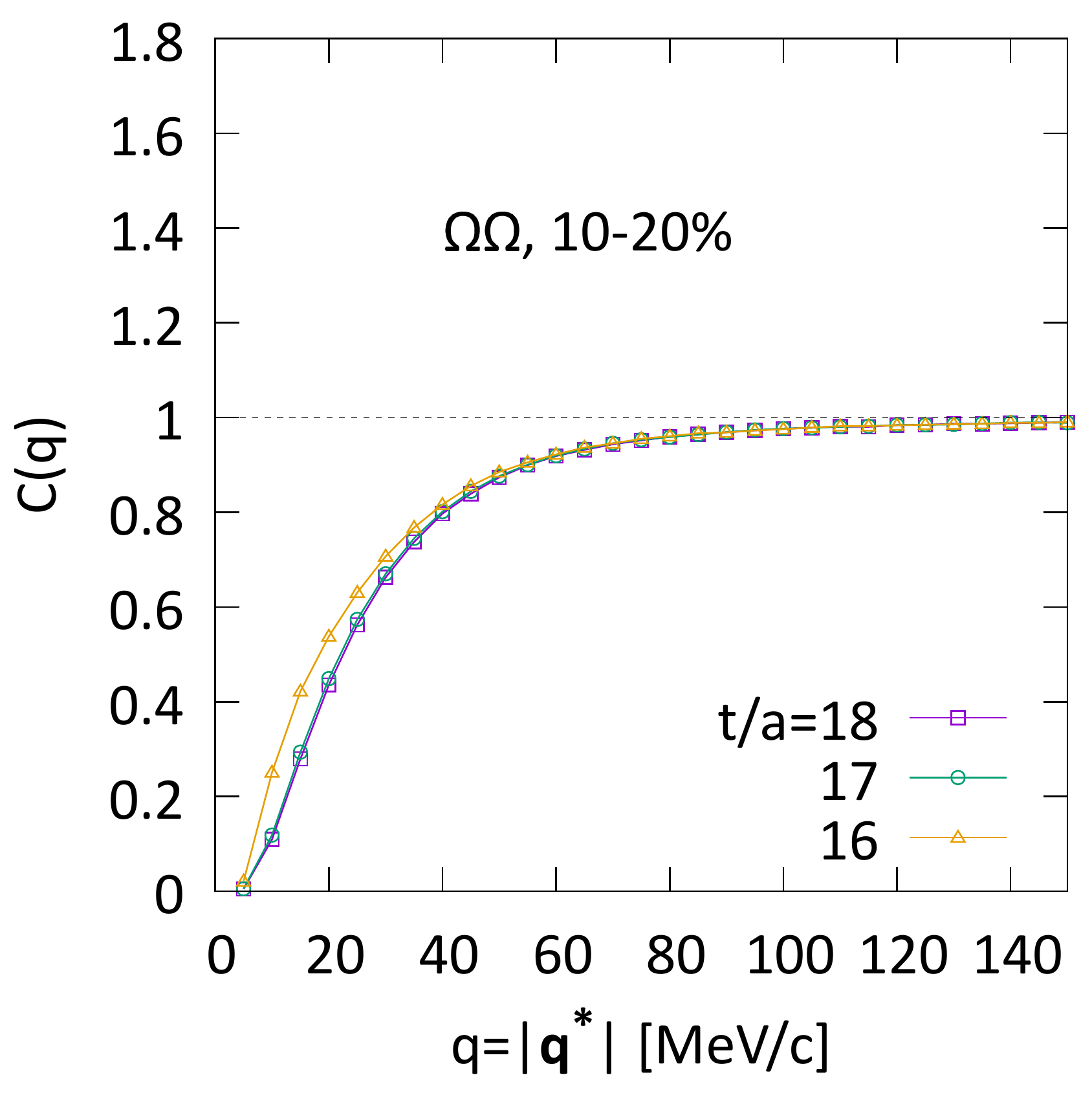}
 \includegraphics[width=0.32\textwidth]{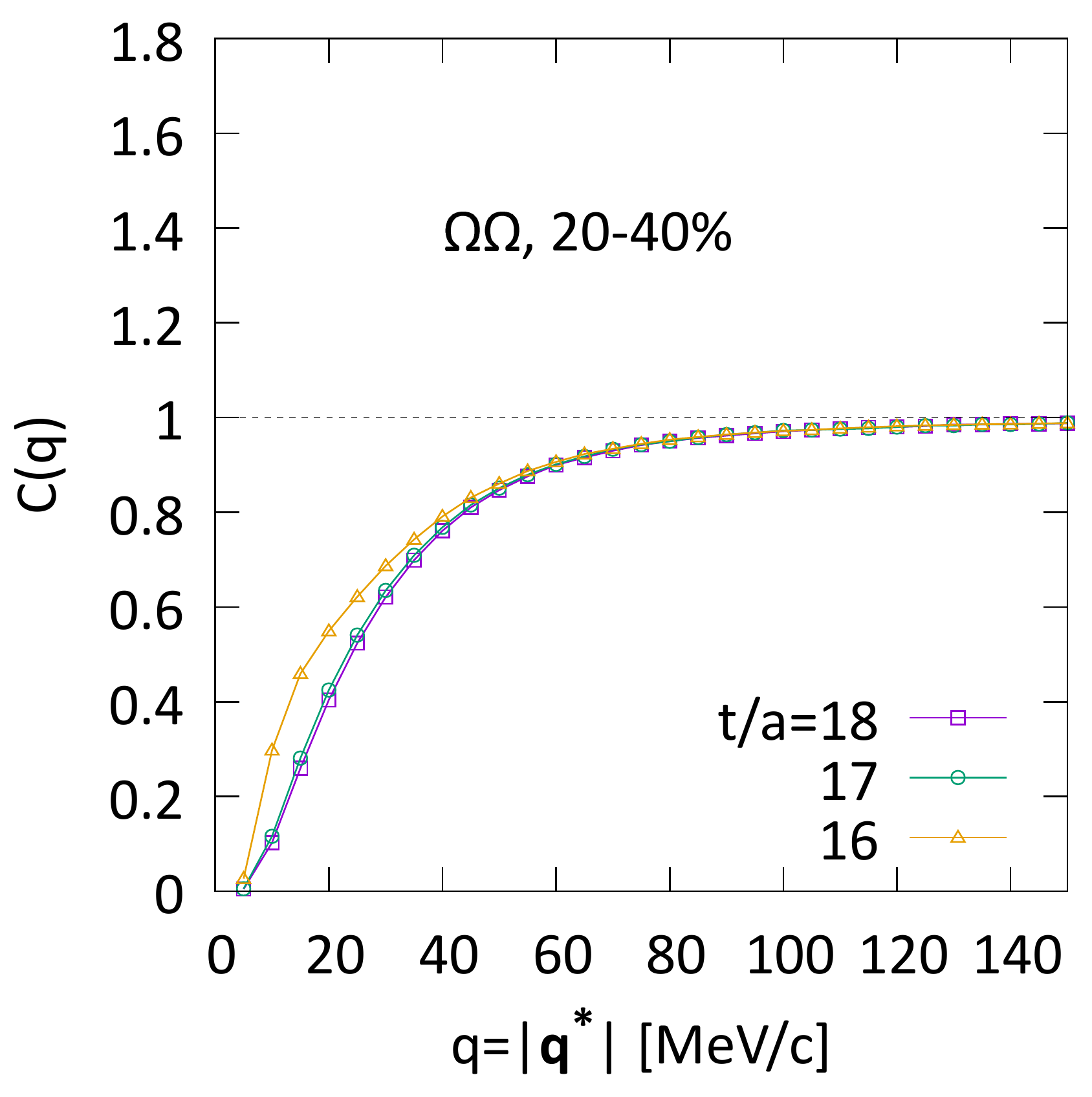}

 \includegraphics[width=0.32\textwidth]{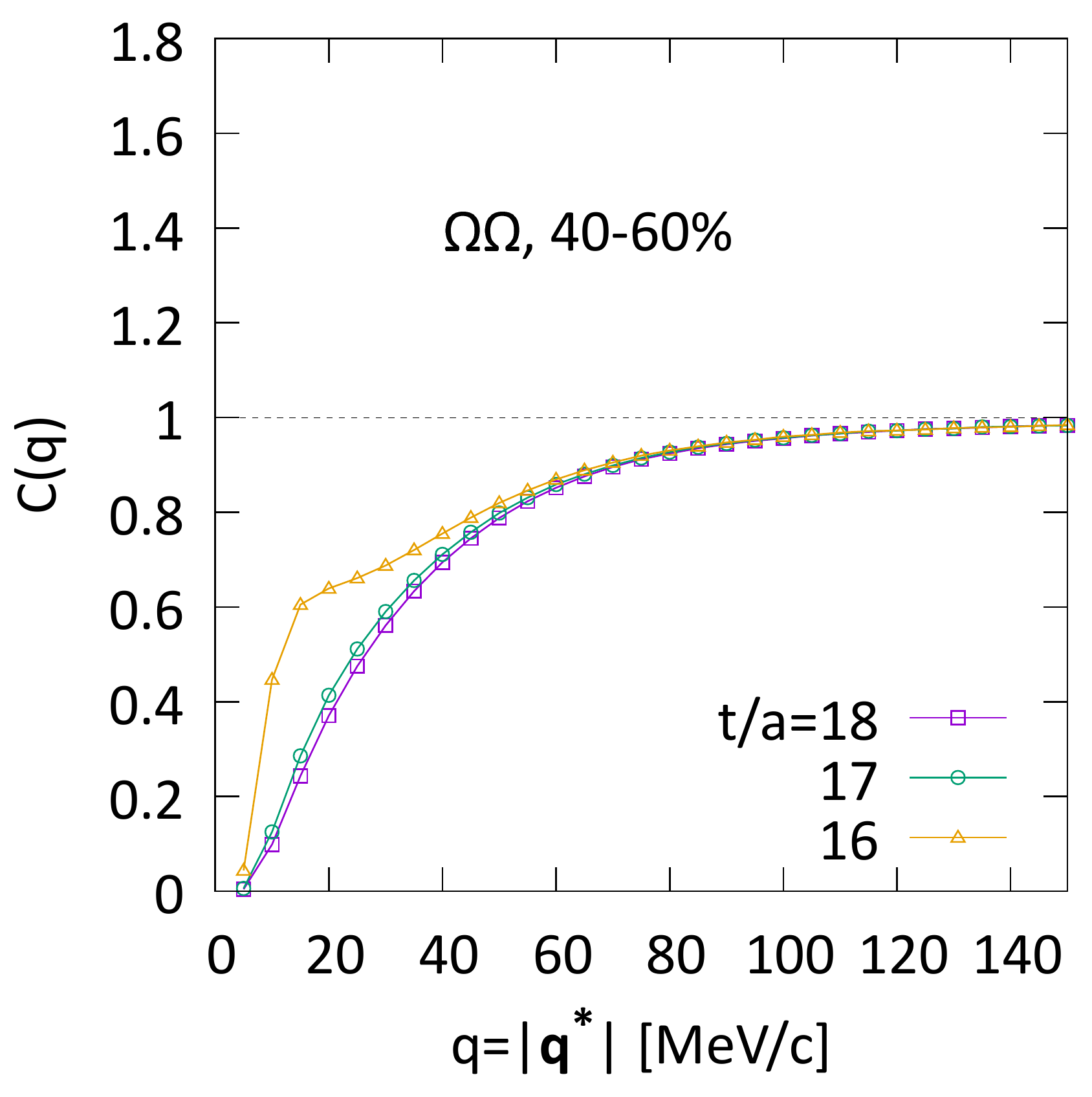}
 \includegraphics[width=0.32\textwidth]{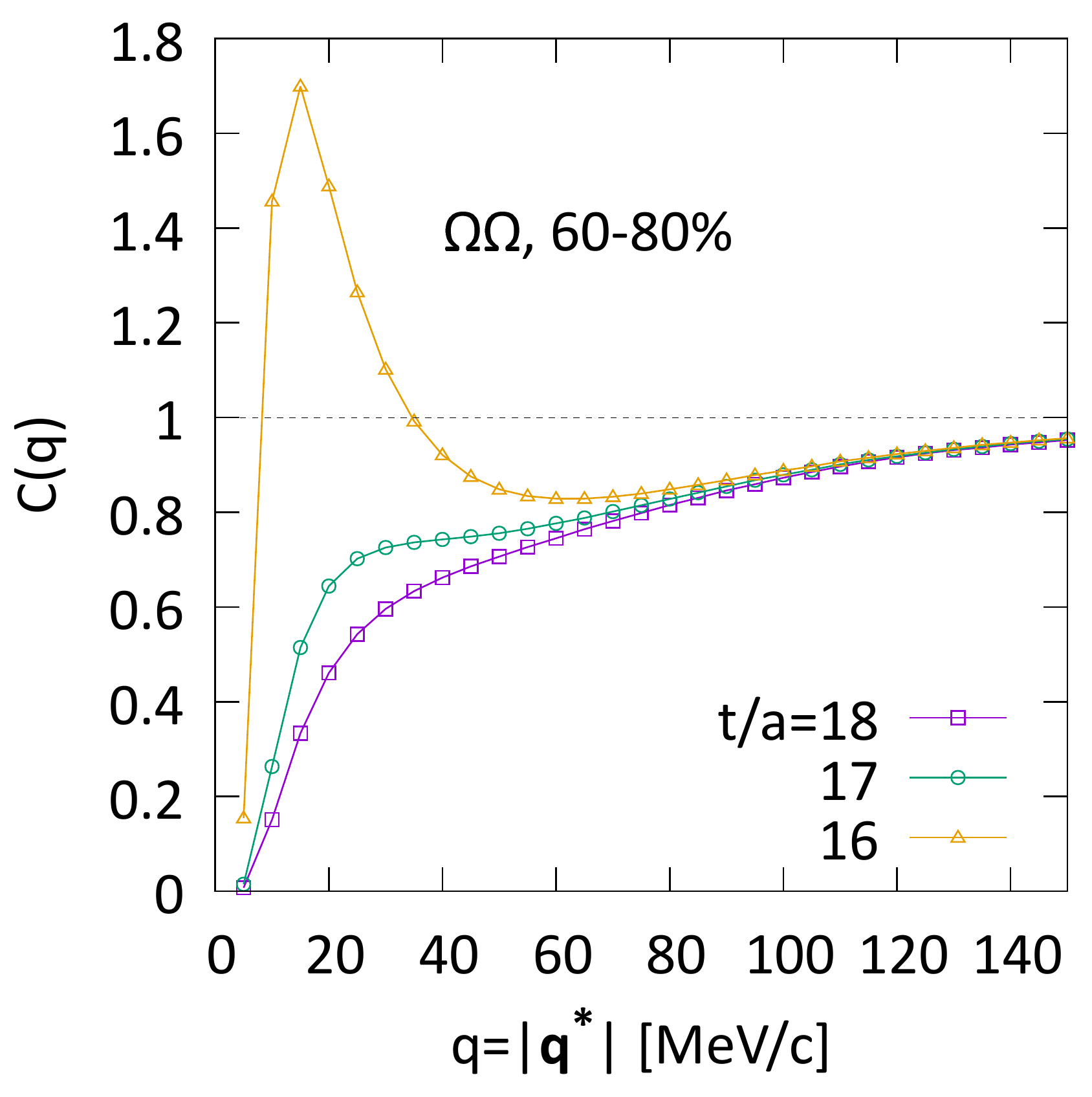}
 \includegraphics[width=0.32\textwidth]{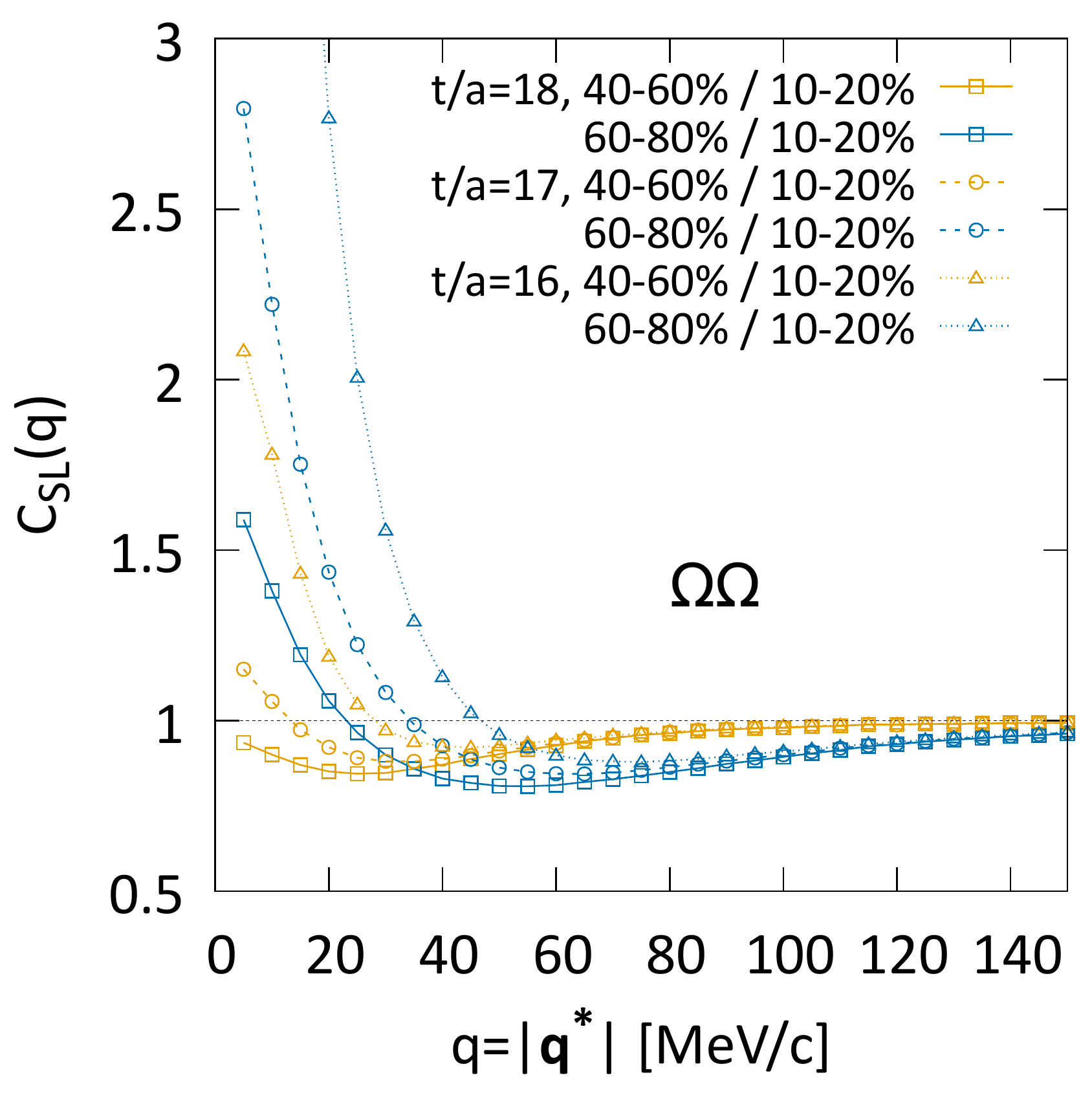}
 \caption{$\Omega\Omega$ correlation function $C(q)$ from central
 (0-10\%)  to peripheral (60-80 \%)  Pb-Pb collisions, as well as  the
 small-to-large ratio $C_\text{SL}(q)$.}
 \label{fig:OmegaOmegaC2}
\end{figure*}

Assuming that the strong interaction except for the $J = 0$ channels is 
negligible, one may write the wave functions \`{a} la Eq.~\eqref{eq:wf_int} 
with the Coulomb repulsion and the Fermi statistics (symmetrization for
$J=0,2$ and anti-symmetrization for $J=1,3$):
\begin{align}
 \varphi^{J=0}(\bm{q},\bm{r}) &=
 \varphi^C_{\text{sym}}(\bm{q},\bm{r})-\varphi^C_{0,\text{sym}}(r) + \chi^C_0(r)
 \label{Eq:pWwf1}\\
 \varphi^{J=2}(\bm{q},\bm{r})&=
 \varphi^C_{\text{sym}}(\bm{q},\bm{r}),  \label{Eq:pWwf2}\\
 \varphi^{J=1,3}(\bm{q},\bm{r})&=
 \varphi^C_{\text{asym}}(\bm{q},\bm{r}).
 \label{Eq:pWwf3}
\end{align}
Here $\varphi^C_{\text{sym}}(\bm{q},\bm{r})$ and
$\varphi^C_{\text{asym}}(\bm{q},\bm{r})$ denote the Coulomb wave
functions with symmetrization and anti-symmetrization, respectively.
Also, $\varphi^C_{0,\text{sym}}(r)$ is the S-wave component of $\varphi^C_{\text{sym}}(\bm{q},\bm{r})$. 
The full wave function in the S-wave, $\chi^C_0(r)$, is obtained by solving
the Schr\"{o}dinger equation
with the strong interaction potential $V_{\rm fit}(r)$
in Fig.~\ref{fig:pot_omega} together with the Coulomb repulsion.
In the absence of the Coulomb interaction, these
expressions reduce to the case of neutral particles, e.g. $\Lambda\Lambda$ pairs
shown in \cite{Morita:2014kza}.
 Also note that the wave functions $\phi^J$
in Eqs.~\eqref{Eq:pWwf1}-\eqref{Eq:pWwf3} contain the higher-partial wave
($L\geq 1$) components.
The total  probability density is thus given by
\begin{equation}
 |\varphi_{\Omega\Omega}(\bm{q},\bm{r})|^2=\sum_{J=0}^{3}\frac{2J+1}{16}|\varphi^J(\bm{q},\bm{r})|^2.
 \label{eq:probJ}
\end{equation}
Note that the effect of the strong interaction in $J=0$ is weighted
only by 1/16 in the probability.

We calculate the correlation function $C(q)$ in Eq.~\eqref{eq:c2_projected}
by combining Eq.~\eqref{eq:expandingsource} and Eq.~\eqref{eq:probJ}.
In the momentum integral, we take vanishing particle rapidities and fix the
transverse momentum to the average values obtained from the spectra
(Fig.~\ref{fig:ptspectra}). In Fig.~\ref{fig:OmegaOmegaC2}, $\Omega\Omega$
correlation functions for different centralities are displayed. 
Note that the system size becomes smaller as the centrality increases.
The depletion of $C(q)$ below 	1  at small $q$  is due to the Coulomb repulsion
and  the HBT effect.  Also, the latter effect extends to wider region of
$q$  for smaller systems. As shown in a schematic analysis given
in Fig.~\ref{fig:LL} (b), the correlation function exhibits stronger FSI
effect 
with decreasing system size. Such a
tendency  can be seen particularly for the $\Omega\Omega$ potential with
$t/a=16$ in Fig.~\ref{fig:OmegaOmegaC2}, since $a_0$ is extremely large.
   
Shown the bottom-right panel of Fig.~\ref{fig:OmegaOmegaC2} is the
small-to-large ratio, $C_\text{SL}(q)$ between 40-60\% (or  60-80\%)
for small systems and 10-20\% for large systems.
Due to the cancellation
of the Coulomb effect, one now finds notable enhancement of
$C_\text{SL}(q)$ above 1 for small $q$ due to the strong $\Omega\Omega$
attraction, and the reduction of $C_\text{SL}(q)$  below 1 for large $q$
due to the HBT effect.

\section{$p\Omega$ correlation}\label{sec:NOmega}

\begin{figure}[h]
 \includegraphics[width=\columnwidth]{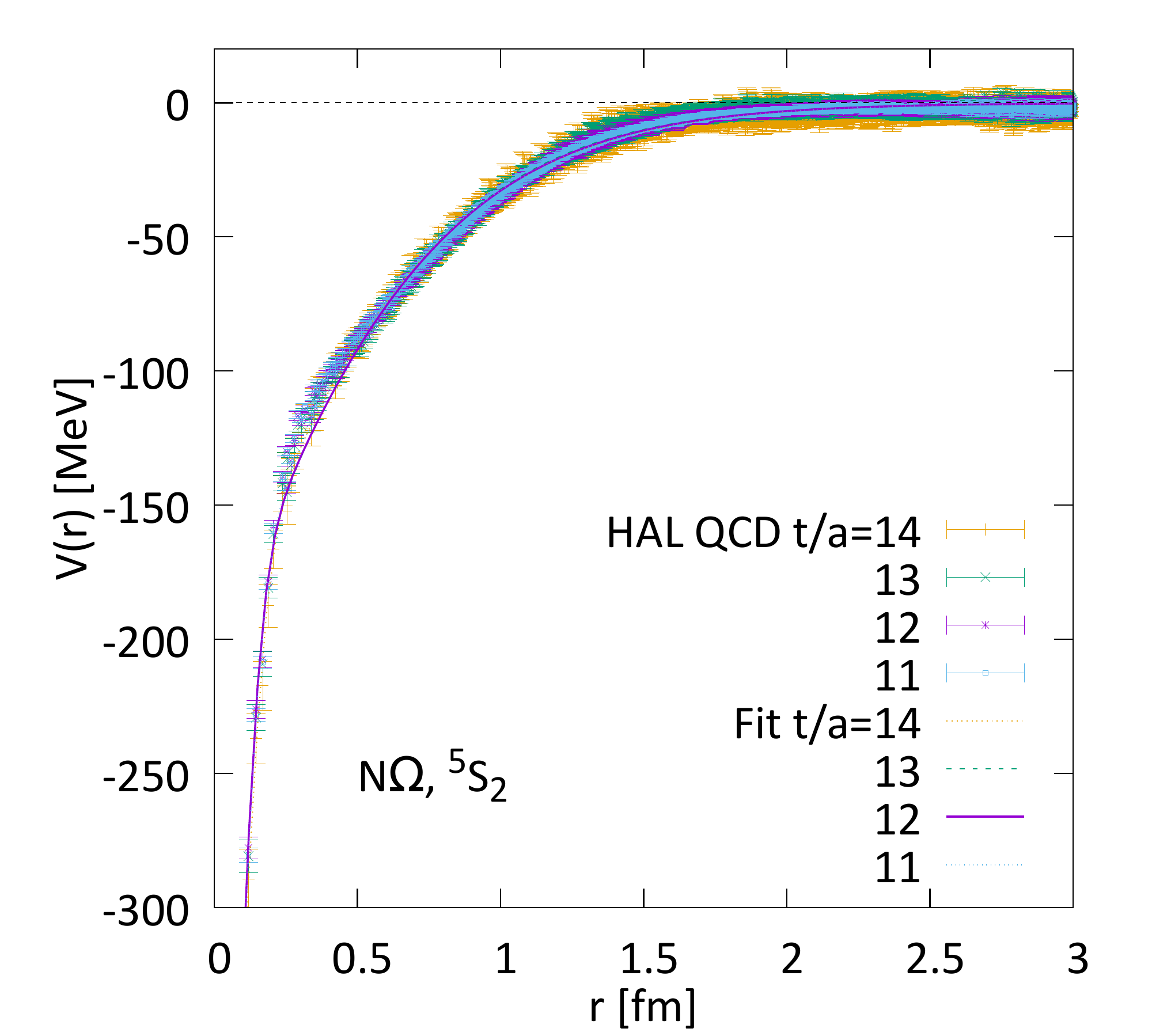} 
 \caption{The S-wave $N\Omega$ potential with $J=2$ from lattice QCD simulations~\cite{Iritani:2018sra}.
 The lattice data are fitted by the form, 
 $V_\text{fit}(r)= b_1 e^{-b_2 r^2} + b_3 (1-e^{-b_4 r^2} ) (e^{-m_{\pi} r}/r)^2$
 with $m_{\pi}$=146 MeV.}
 \label{fig:nomegapot}
\end{figure}

Let us now move on to the  results for $p\Omega$ correlations.
Among {$J=1\ (^5S_2)$ and $J=2\ (^3S_1)$ channels
which the $p\Omega$ pair can take,
the $J=2$ channel is expected to have a shallow bound state as indicated from lattice QCD
 \cite{Iritani:2018sra}. 
Note, however, that the $p\Omega$ pair is not the lowest energy channel
in the $S=-3$ dibaryon system:
There exist thresholds of the octet-octet states ($\Lambda \Xi$ and
$\Sigma \Xi$) at lower energies,  which act as absorptive channels for
$p\Omega$.  The S-wave $J=2$ channel couples to octet-octet states
only through the $D$ wave, so that  the decay is dynamically suppressed
and its effect on the correlation function is considered to be
sufficiently small. According to Ref.~\cite{Sekihara:2018tsb}, where the 
 $J=2$ $N\Omega$ interaction is discussed with the meson exchange model
including the decay channels,
the coupling does not change the weak-binding nature of $p\Omega$.
Thus, in the following calculations, we apply the single-channel
approximation to the $J=2$ $p\Omega$ correlation function.

In the previous study
on $C_\text{SL}(q)$ for $p\Omega$~\cite{Morita:2016auo},
the $J=2$ potential obtained by lattice QCD simulations with heavy quark
masses \cite{Etminan:2014tya} were used.
Below, we update the analysis by using the $J=2$ potential for  nearly
physical quark masses as described below.

\subsection{$N\Omega$ interaction from lattice QCD}
  
The $N\Omega$ interaction in $J=2$ channel has been calculated  by
(2+1)-flavor lattice QCD simulations~\cite{Iritani:2018sra} with the same
setup as the $\Omega\Omega$ case discussed in Sec.\ref{subsec:OO-L}.
In this case, the  Euclidean time interval was  chosen to be $t/a=11-14$
to avoid significant statistical errors for large $t$.
Resultant potentials  with statistical errors are recapitulated in
Fig.~\ref{fig:nomegapot} together with the fitted potential of a
Gaussian + (Yukawa)$^2$ form. The scattering length and the effective
range without the Coulomb interaction are $a_0 \simeq $ 5.3 fm and
$r_\text{eff} \simeq$ 1.26 fm, respectively,  so that a weakly bound
$N\Omega$ appears with the binding energy  $E_B \sim $ 1.54 MeV.
  
Table \ref{tbl:NW_scattering} shows the low energy scattering parameters and binding energies
obtained by solving the Schr\"{o}dinger equation in the presence of  the
attraction from the strong interaction and the extra  attraction from the Coulomb interaction.
The value of the resultant scattering length is compatible with the
expected effective system size in heavy-ion collisions, thus one can
expect characteristic depletion of the correlation function and its
variation for the system with bound state, against system size
as seen from Fig.~\ref{fig:LL}.}

\begin{table}[h]
 \caption{S-wave scattering length $a_0$, effective range $r_{\text{eff}}$, and
 binding energy of the $p\Omega$ pair with the lattice QCD potential for different $t/a$ and 
 the Coulomb attraction.}
 \label{tbl:NW_scattering}
 \begin{tabular}[t]{c|ccc}\hline
  $t/a$ & $a_0$ [fm] & $r_{\text{eff}}$ [fm] & $E_B$ [MeV]\\\hline
  11    & 3.45 & 1.33 & 2.15 \\
  12    & 3.38 & 1.31 & 2.27 \\
  13    & 3.49 & 1.31 & 2.08 \\
  14    & 3.40 & 1.33 & 2.24 \\ \hline
 \end{tabular}
\end{table}

\subsection{Correlation function}

\begin{figure}[t]
 \centering
 \includegraphics[width=\columnwidth]{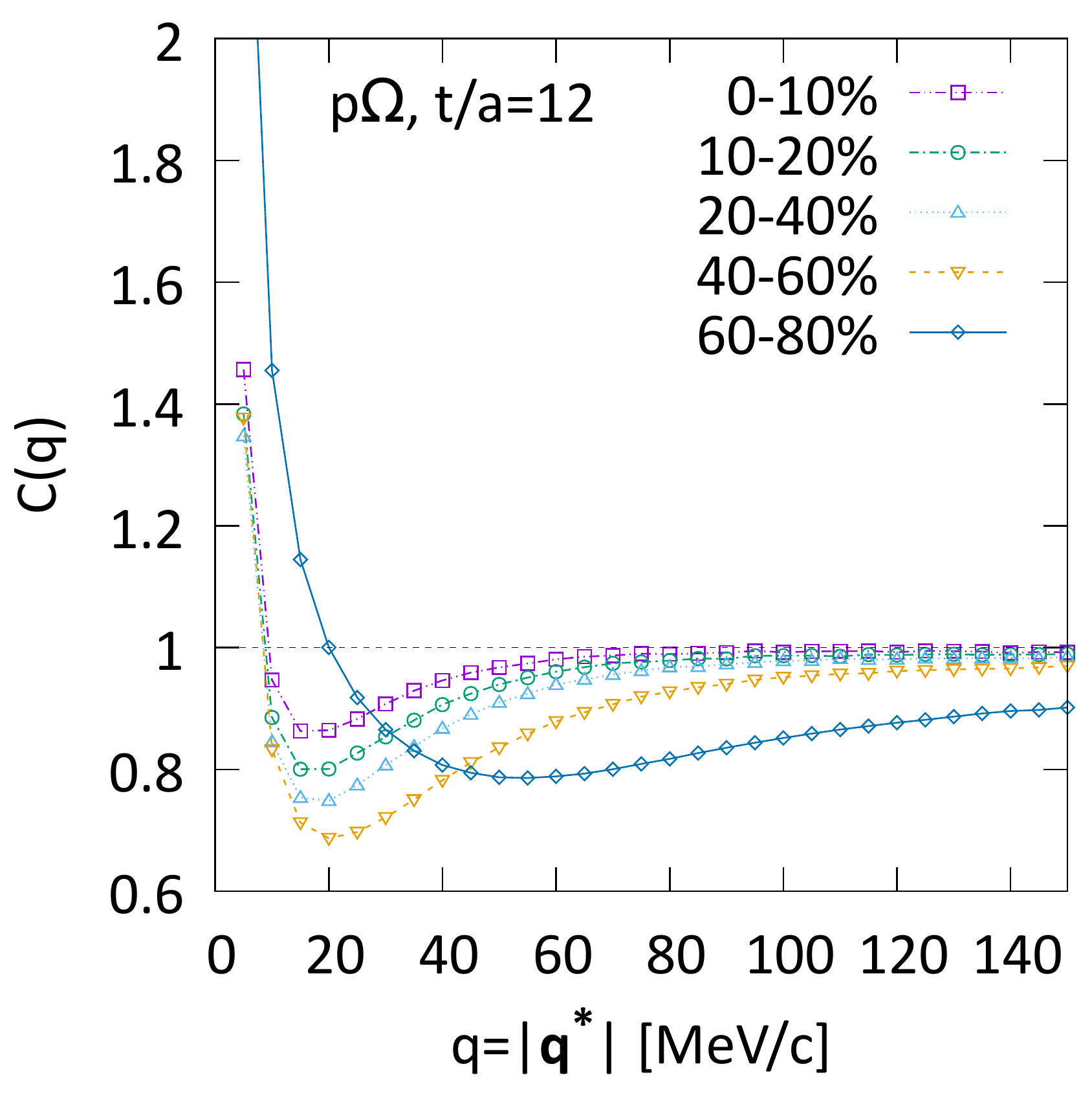}

 \includegraphics[width=\columnwidth]{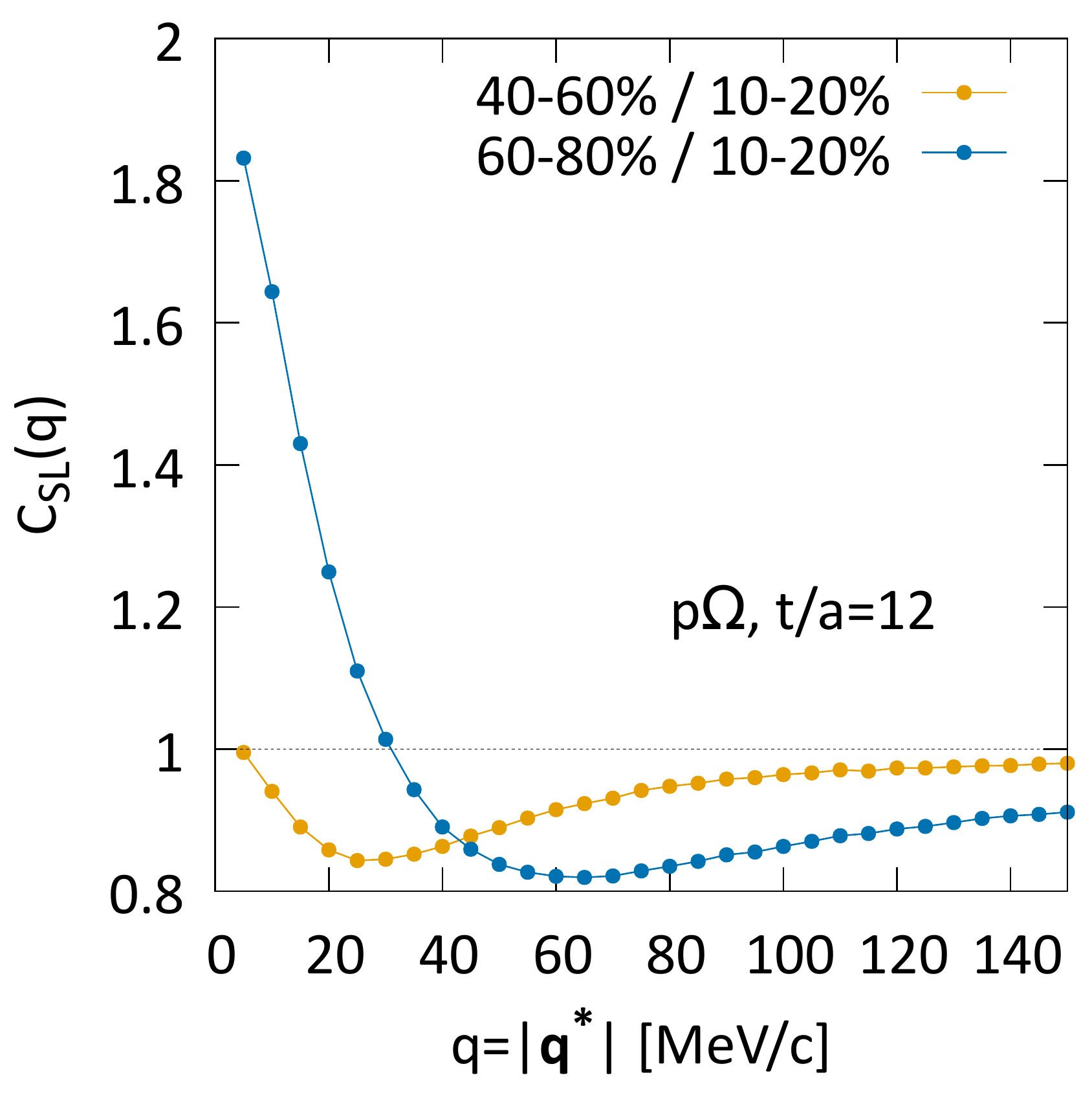}

 \caption{$p\Omega$ correlation function from central (0-10\%)  to peripheral (60-80 \%) 
 Pb-Pb collisions (upper panel), as well as from peripheral to central
 collisions and the small-to-large ratio (lower panel).}
 \label{fig:c2_p-omega}
\end{figure}

In addition to the $J=2$ channel,
the $N\Omega$ system has the $J=1$ channel 
which is expected to couple strongly with low-lying octet-octet states
due to fall apart decay in the S-wave.
In the same way as Ref.~\cite{Morita:2016auo}, we consider a limiting
case where the $J=1$ $p\Omega$ pairs are perfectly absorbed into
low-lying states  through the potential  $V^{J=1}(r) = -i\theta(r_0-r)V_0$.
The strength $V_0$ is taken to be infinity and $r_0$ is set to 2 fm 
where Coulomb interaction dominates  over the $J=1$ LQCD potential.
Accordingly, the wave function is  written as
$\varphi^{J}(\bm{q},\bm{r})=\varphi^C(\bm{q},\bm{r})-\varphi^C_0(r)+\chi_0^C(r)$,
where the  scattering wave function in the S-wave, $\chi_0^C(r)$,
receives the effects of the interactions.

Then the total probability density reads
\begin{equation}
 |\varphi_{p\Omega}(\bm{q},\bm{r})|^2 = \sum_{J=1}^{2}\frac{2J+1}{8}|\varphi^J(\bm{q},\bm{r})|^2.
\end{equation}
Here the $J=2$ contribution which is of our interest, is weighted by a large factor $5/8$.
The number of the low momentum pairs decrease due to the absorption in the $J=1$ channel
and the resultant correlation function $C(q)$  tends to  decrease but not with significant amount as 
discussed in in Ref.~\cite{Morita:2016auo}.

Figure \ref{fig:c2_p-omega} shows the $p\Omega$ correlation functions
from peripheral to central collisions.  Since the $N\Omega$ potential in
Fig.~\ref{fig:nomegapot} is  nearly independent of $t/a$,  the same 
holds for $C(q^*)$ too. Thus we display only results of $t/a=12$.
The enhancement of $C(q)$ above 1 for small $q$ is due to the Coulomb
attraction whereas the suppression of $C(q)$ below 1  is due to the
positive scattering length, or equivalently the existence of $p\Omega$
bound state. The effect of FSI is  smallest (largest) in central
collisions $(0-10 \%)$ (peripheral collisions $(60-80 \%)$),
so that the region of the suppressed correlation becomes deeper and wider
as the system size decreases, in accordance with the moderate value of the
scattering length ($a_0 \simeq 3.4 $ fm)  in Table \ref{tbl:NW_scattering}.

Shown the bottom panel of Fig.\ref{fig:c2_p-omega} is the
small-to-large ratio, $C_\text{SL}(q)$, between 40-60\% (or  60-80\%) for
 the small system and 10-20\% for the large system.
After the cancellation of the Coulomb effect,
one now finds notable enhancement
of $C_\text{SL}(q)$ above 1 at small $q$
 and depletion below 1 at $q =(20-80)~\mathrm{MeV}$
due to the strong $p\Omega$ attraction accommodating a bound state.
In response to the theoretical proposal 
in \cite{Morita:2016auo},
the STAR collaboration at RHIC has reported a first measurement
of $p\Omega$ correlation in Au+Au collisions~\cite{STAR:2018uho}.
Although the statistics of the data are not sufficient to draw a definitive
conclusion, the measured  $C(q)$ and $C_\text{SL} (q)$ 
show similar tendency with  Fig.\ref{fig:c2_p-omega} in the present paper.

\section{Summary and Concluding remarks}
\label{sec:conclusion}

We have studied the  two-particle momentum correlations for
$\Omega\Omega$ and $p\Omega$  in relativistic heavy-ion collisions.
The correlation functions are calculated by using an expanding source model 
combined with the latest lattice QCD potentials which predict  shallow
bound states with  relatively large positive scattering lengths
in the $J=0$ $\Omega\Omega$ and the $J=2$ $N\Omega$.

At the LHC energies, the correlation function $C(q)$  for
$\Omega\Omega$  in Pb-Pb collisions exhibits an enhancement due to large
scattering length ($a_0 >$ 10 fm)  over the Coulomb repulsion 
and the  HBT effect, especially in the peripheral
events.  This characteristic feature can be best visible and quantified 
as an enhancement of the small-to-large ratio  $C_{\text{SL}}(q)$ at  $q < 40 $ MeV/c.

On the other hand, the characteristic feature of the  correlation
function $C(q)$ of $p\Omega$  is its depletion below 1 at $q = 20-40 $
MeV due to the moderately large value of the positive scattering length $a_0 \simeq 3.4$ fm.
Properly chosen  small-to-large ratio  $C_{\text{SL}}(q^*)$ also exhibits this behavior.

Measuring the $\Omega\Omega$ in heavy-ion collisions is a challenge
even with the high luminosity upgrade of LHC due to its small production rate
as well as the correlation measurement at small $q \ (< 50$ MeV).
Therefore, not only the luminosity upgrade but also the improvements
of measurement techniques would be necessary.
 
In response to our theoretical proposal in \cite{Morita:2016auo},
the STAR collaboration at RHIC has reported a first measurement of
$p\Omega$ correlation in Au+Au collisions~\cite{STAR:2018uho}.
Although the statistics of the data are not sufficient
to draw a definitive conclusion, the measured $C(q)$ and $C_\text{SL} (q)$ 
show similar tendency with Fig.\ref{fig:c2_p-omega} in the present paper. 
Also the ALICE Collaboration at LHC has started the $p\Omega$ measurements
with $pp$ and $p$-Pb collisions ~\cite{ALICE:pOmega}.
Extracting the $p\Omega$ interaction from a combined theoretical analysis 
of the $pp$, $pA$ and $AA$ collisions with proper uncertainty quantification 
would be an interesting future problem.
(See Appendix A for an exploratory study along such direction.) 

In order to draw definite conclusion on the existence of the $\Omega\Omega$
and $N\Omega$ dibaryon bound states from the future and existing
correlation function data, we need further works to be done.
First, it is desired to obtain
not only the $J=0$ $\Omega\Omega$ potential and $J=1$ $N\Omega$ potential
but also the $J=1,2$ and $3$ $\Omega\Omega$ potentials
and the $J=1$ $N\Omega$ potential. 
Second, the coupled channel effects need to be clarified.
As discussed in the Appendix A, the $J=1$ contribution causes
visible uncertainties in the $p\Omega$ correlation function. 
While the coupling effects to octet-octet channels
with $J=1$ in the $p\Omega$ correlation function
have been assumed to be described by the absorption,
the coupled channel formula~\cite{Haidenbauer:2018jvl} shows that
creation processes such as $\Lambda\Xi \to p\Omega$ also contribute
to the correlation function of $p\Omega$.
Then we need to evaluate the transition potentials and the source function
of $\Lambda$ and $\Sigma$.

\section*{Acknowledgments}
The authors thank
Takumi Iritani and Takumi Doi 
for useful discussions and valuable help in preparing the manuscript.
The authors also thank
Sinya Aoki, Kenji Sasaki, 
Neha Shah, 
Laura Fabbietti, 
Valentina Mantovani Sarti, 
Ot\'on V\'azquez Doce, 
Johann Haidenbauer, 
and other participants of the YITP workshop (YITP-T-18-07)
for useful discussions.
This work is supported in part by the Grants-in-Aid for Scientific Research
from JSPS (Nos.
19H05151, 
19H05150, 
19H01898, 
18H05236, 
and 16K17694), 
by the Yukawa International Program for Quark-hadron Sciences (YIPQS)
by the Polish National Science Center NCN under Maestro grant
EC-2013/10/A/ST2/00106, 
by the National Natural Science Foundation of China (NSFC)
and the Deutsche Forschungsgemeinschaft (DFG) through the funds
provided to the Sino-German Collaborative Research Center
``Symmetries and the Emergence of Structure in QCD"
(NSFC Grant No. 11621131001, DFG Grant No. TRR110),
by the NSFC under Grant No. 11747601 and No. 11835015,
and
by the Chinese Academy of Sciences (CAS)
under Grant No. QYZDB-SSW-SYS013 and No. XDPB09.

\appendix

\section{System size dependence of $C(q)$ for $p\Omega$
 with uncertainty quantification }

\begin{figure*}[t]
 \includegraphics[width=0.32\textwidth]{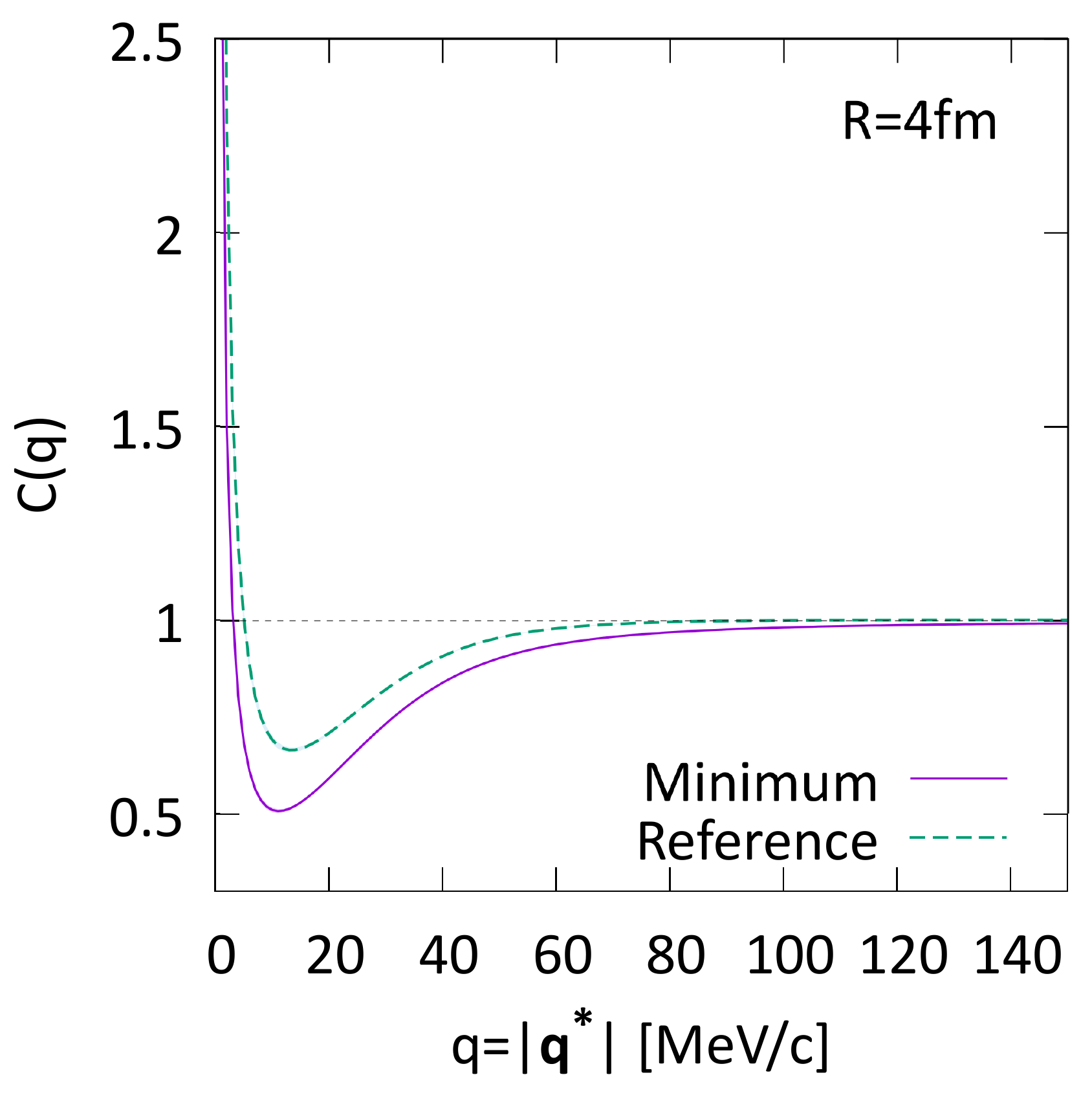}
 \includegraphics[width=0.32\textwidth]{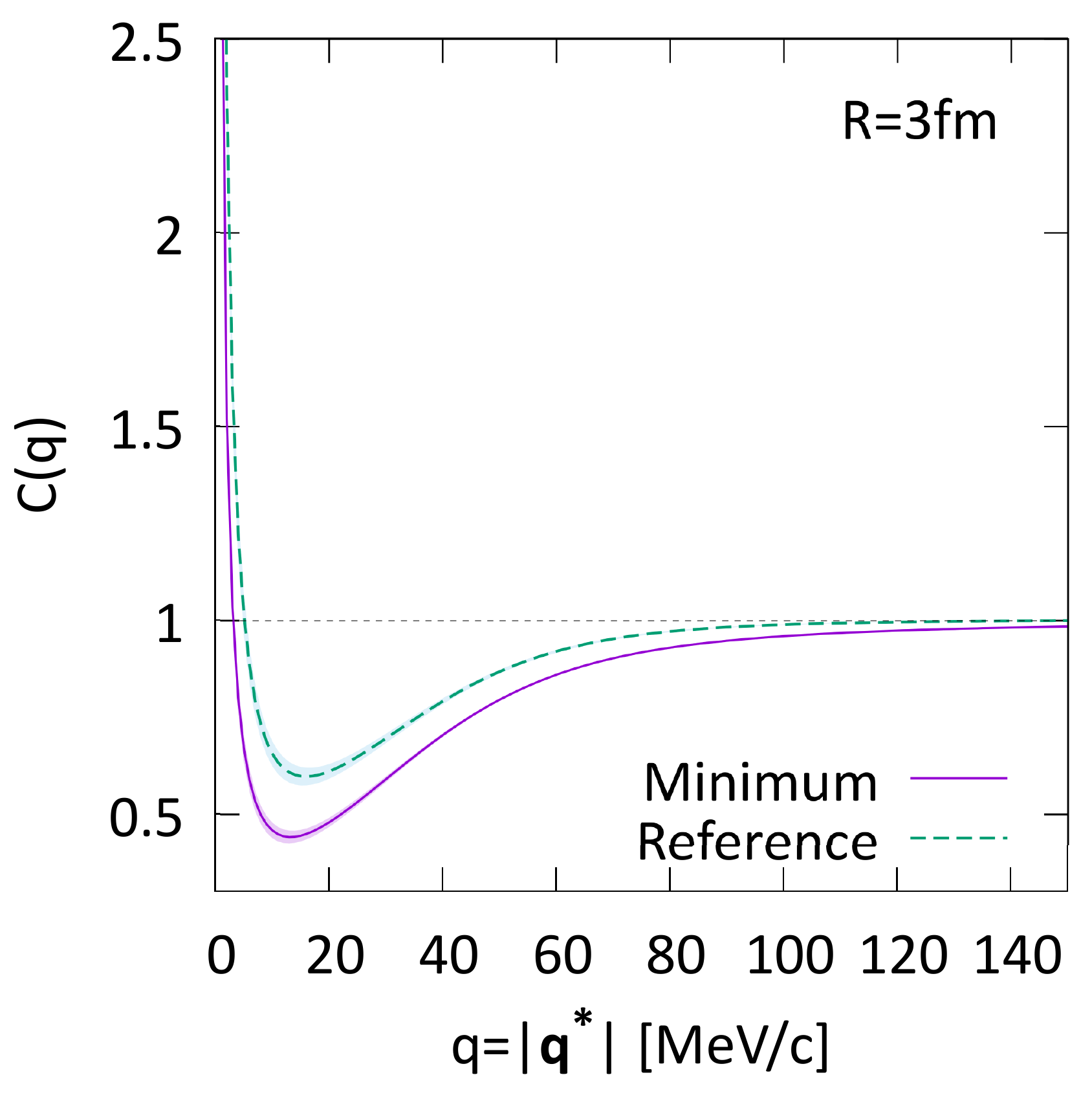}
 \includegraphics[width=0.32\textwidth]{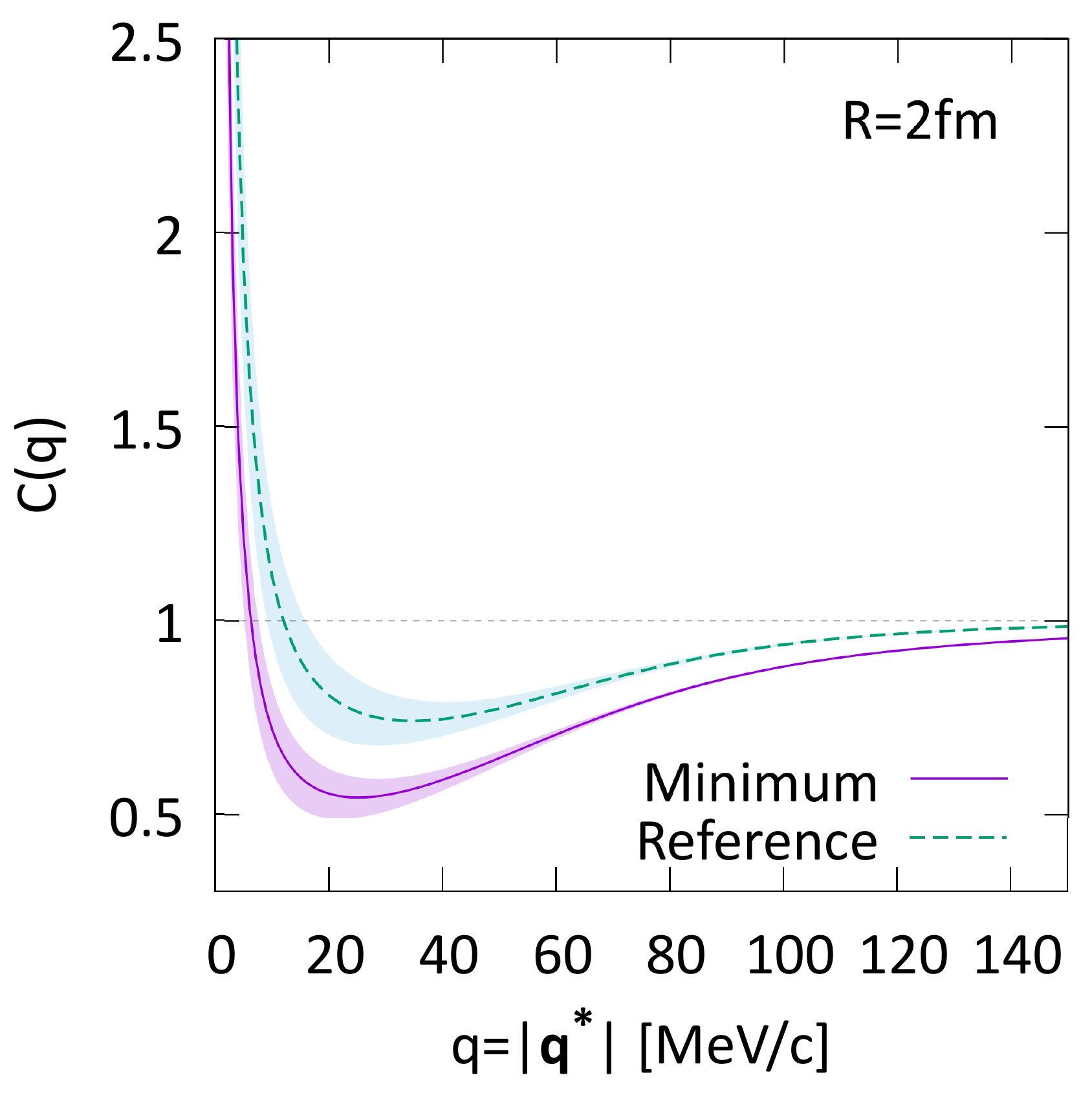}
 
 \includegraphics[width=0.32\textwidth]{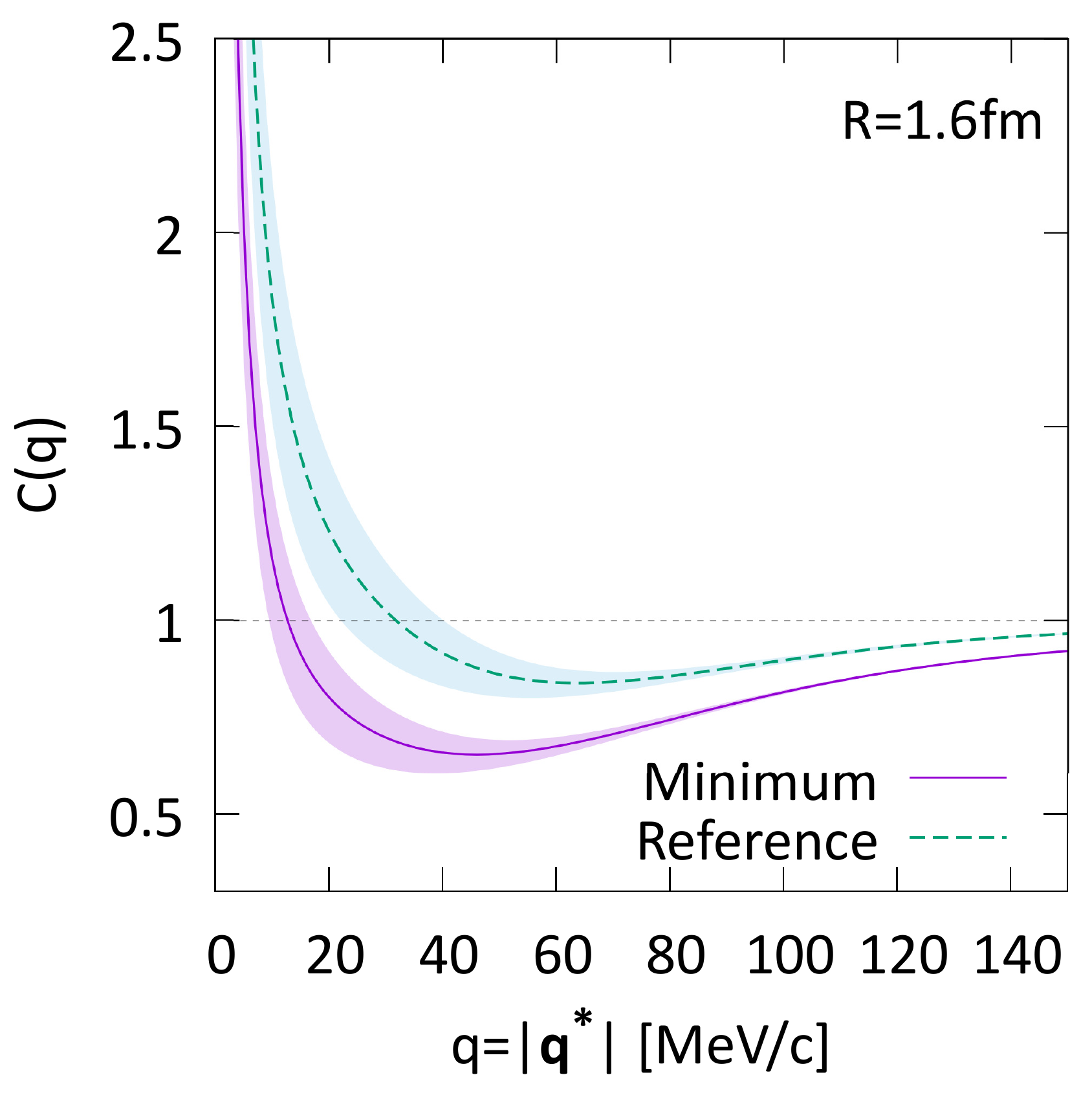}
 \includegraphics[width=0.32\textwidth]{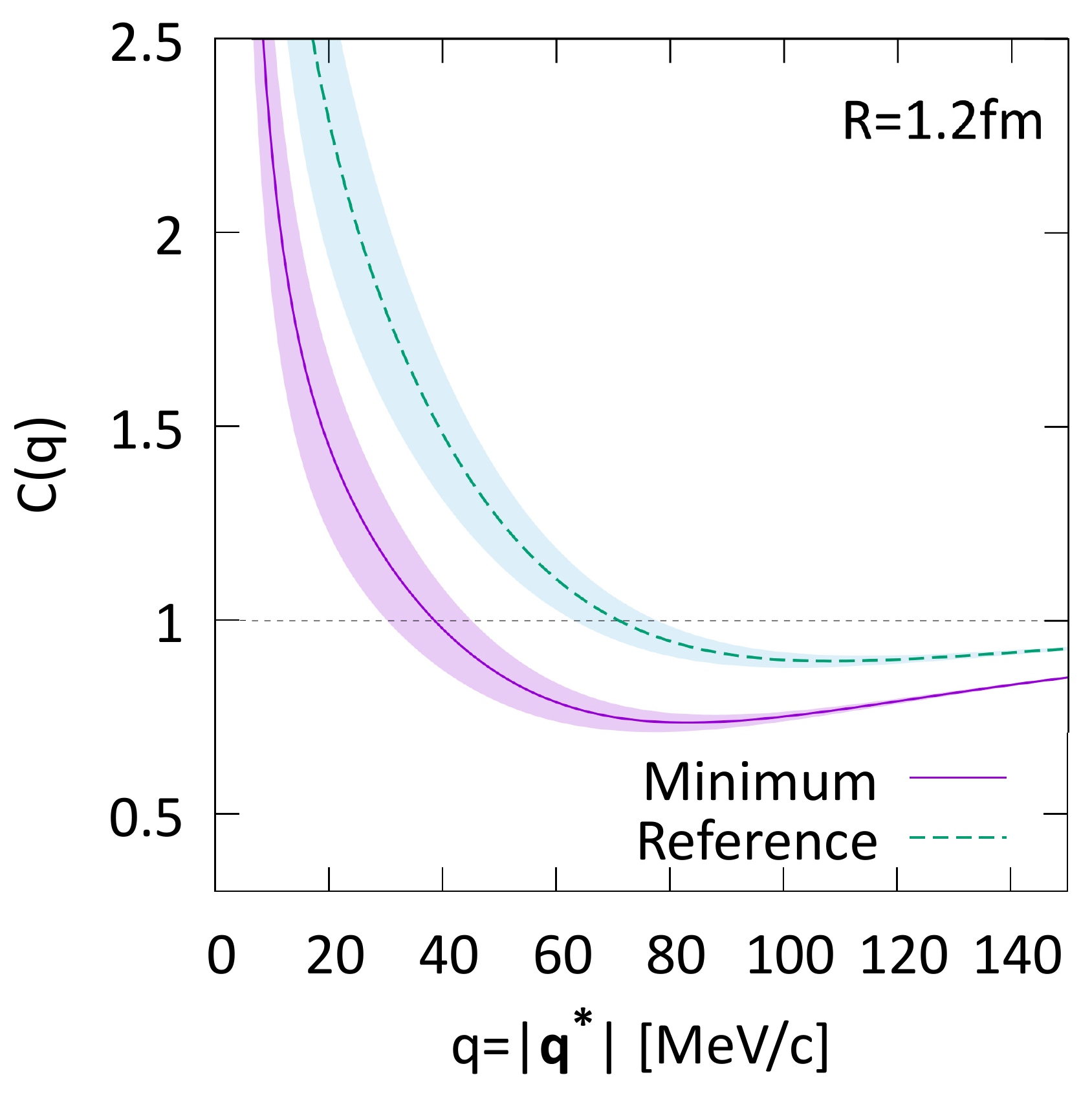}
 \includegraphics[width=0.32\textwidth]{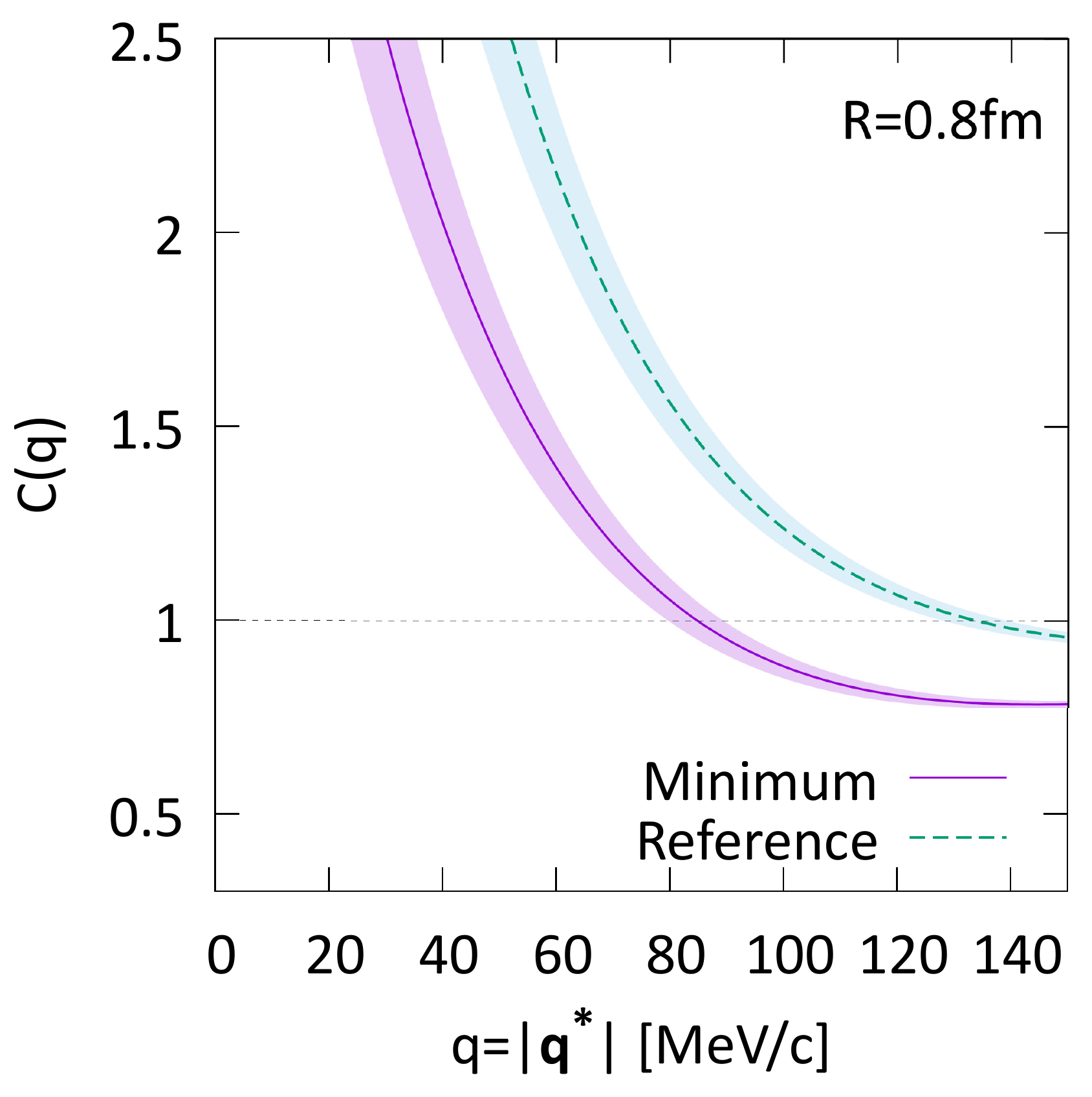}
 \caption{$p\Omega$ correlation function calculated with the static
 Gaussian source function employing the $t/a =12$ potential. The purple
 solid line denotes the result with $\chi_0^{C,J=1} = 0$, and blue
 dashed line denotes the result with the assumption of
 $\chi_0^{C,J=1}=\chi_0^{C,J=2}$. Gaussian source size is chosen to be
 in the range $R=0.8$ - $4\ \mathrm{fm}$. The error of each correlation
 estimated with the Jackknife method is shown by the colored shadow.}
	\label{fig:p-omega_Jerror}
\end{figure*}

In the light of feasibility of measuring $p\Omega$
correlation in $pp$, $pA$ and $AA$ collisions, it is desirable to
get a feel for theoretical uncertainties in evaluating the momentum correlations. 
In the following,
we focus on the uncertainties originating from the $J=2$ $p\Omega$
potential from  lattice QCD and from the treatment of the unknown $J=1$
$p\Omega$ potential. 
To make the discussion transparent, we consider a simplified static and
spherically symmetric Gaussian source
function $S(r)=(4\pi R^2)^{-3/2}\exp(-r^2/4R^2)$ 
with the source size ranging from 0.8 to 4 fm. 

For uncertainties arising from the insufficient information 
on the  $J=1$ potential, 
we evaluate  ``minimum'' and  ``reference'' contributions
from the  $J=1$ channel. The ``minimum''  is obtained 
by assuming $\chi^{C, J=1}_0(r) = 0$, i.e. complete absorption of the wave function in all range of $r$.
 This leads to the minimum value of $C(q)$ as seen
in Eq.~\eqref{eq:corr_static}.
The ``reference'' is obtained by assuming 
$\chi^{C,J=1}_0(r) = \chi^{C,J=2}_0(r)$, i.e.
the same attraction between $J=1$ and $J=2$ without absorption.
The statistical uncertainty for each case is estimated
by the statistical error of the $J=2$
$N\Omega$ potential at $t/a=12$
 by the Jackknife method in the similar way as ~\cite{Iritani:2018sra}.

The results of $C(q)$ for different values of $R$ are shown in Fig.~\ref{fig:p-omega_Jerror}.
The shaded areas represent the statistical errors 
 obtained from the Jackknife analysis.
For  $R \leq 2 $ fm, the ``minimum'' and ``reference''
correlation functions
exhibit sizable  differences with larger statistical uncertainty.
 This is because the condition for the unitary region
  shown in Fig.~\ref{fig:LL} begins to hold
 with $a_{0}\simeq 3.4$ fm in Table~\ref{tbl:NW_scattering}),
 so that the correlation function becomes more sensitive to the 
 uncertainty of the potential as well as the treatment of the $J=1$ channel.

Within the above uncertainty estimate, we can safely conclude that 
the correlation function can be strongly suppressed 
at $q < 40~\mathrm{MeV}$ for systems with $2\text{ fm}\lesssim R\lesssim 4$ fm.
We also find that the suppressed region of $C(q)$ moves
toward the lower $q$ direction with increasing source size.
This behavior is consistent with the trend found in the data from
Au+Au collisions by the STAR Collaboration at RHIC~\cite{STAR:2018uho}.
By comparison, strong enhancement at small momenta would be observed 
for small systems with $R \simeq 1~\mathrm{fm}$
as found in the preliminary data
by the ALICE Collaboration at LHC~\cite{ALICE:pOmega}.

\section{Comparison of $N\Omega$ potentials}

\begin{figure}[h]
 \includegraphics[width=\columnwidth]{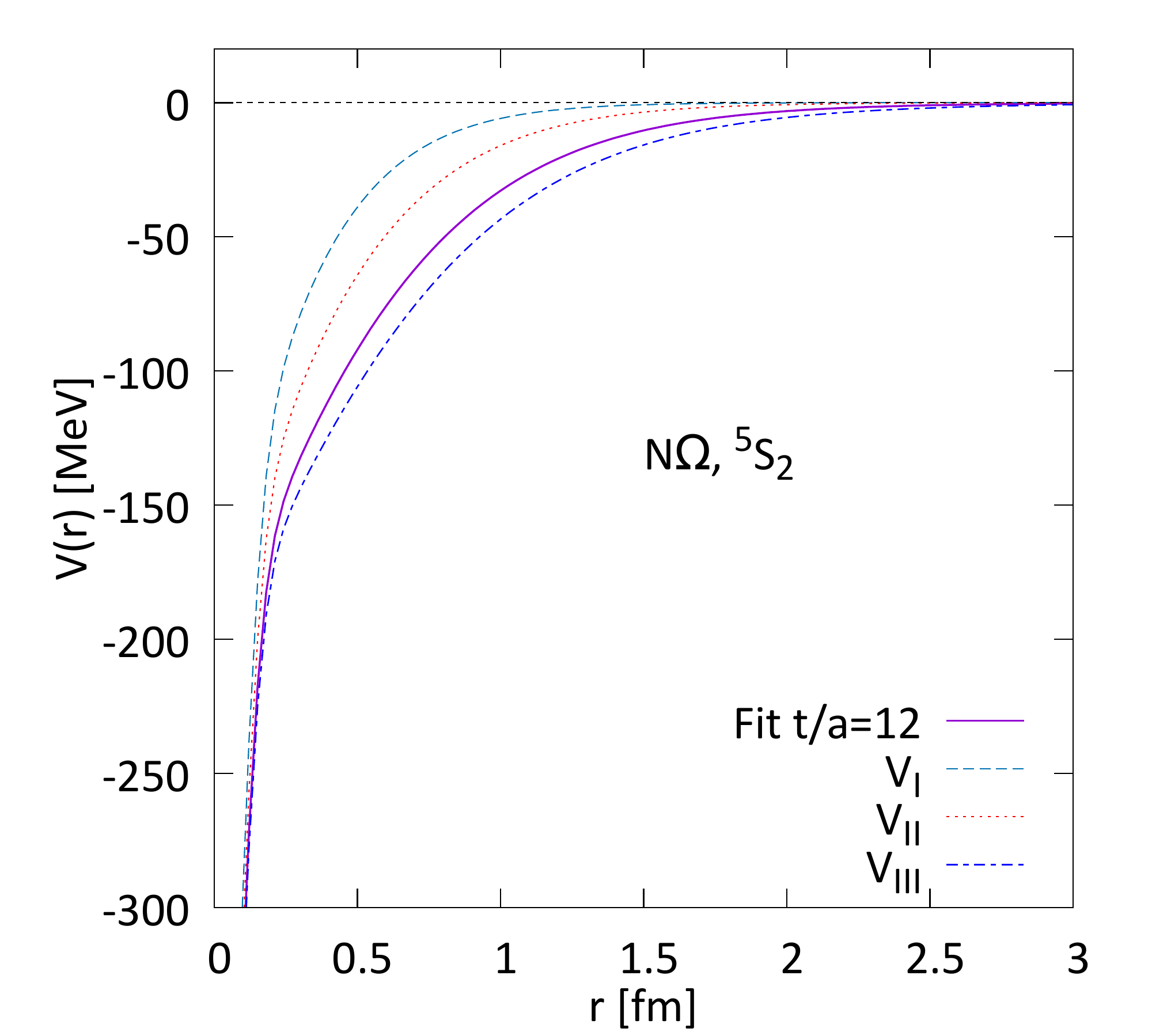}
 \caption{Comparison of the S-wave $N\Omega$ potentials with $J=2$
 in Refs.~\cite{Iritani:2018sra} and \cite{Morita:2016auo}.
 The solid curve show the $N\Omega$ potential with nearly physical quark
 masses~\cite{Iritani:2018sra} at $t/a=12$.
 The dashed, dotted, and dash-dotted curves show 
 the $J=2$ $N\Omega$ potentials, $V_\mathrm{I}$, $V_\mathrm{II}$
 and $V_\mathrm{III}$, given in \cite{Morita:2016auo}.}
 \label{fig:NW_comparison}
\end{figure}

We here compare the $J=2$ $N\Omega$ potential
used in this work and those used in \cite{Morita:2016auo}.
The former is obtained from LQCD simulations with nearly physical quark masses
($m_\pi=146~\mathrm{MeV}$ and $m_K=525~\mathrm{MeV}$) ~\cite{Iritani:2018sra}, 
while the latter is those with heavier quark masses
($m_\pi=875~\mathrm{MeV}$ and $m_K=916~\mathrm{MeV}$)~\cite{Etminan:2014tya}.
In Fig.~\ref{fig:NW_comparison}, we show the $J=2$ $N\Omega$ potential
with nearly physical quark masses at $t/a=12$ (solid curve),
and the potentials given in \cite{Morita:2016auo},
$V_\mathrm{I}$ (dashed), $V_\mathrm{II}$ (dotted)
and $V_\mathrm{III}$ (dash-dotted).
The potential $V_\mathrm{II}$ is the best fit of the lattice data
with heavier quark masses with a form 
$V_\text{fit}(r)= b_1 e^{-b_2 r^2} + b_3 (1-e^{-b_4 r^2} ) (e^{-b_5 r}/r)^2$.
$V_\mathrm{I}$ and $V_\mathrm{III}$ are two typical examples
with weaker and stronger attractions, respectively.
These potentials together with the Coulomb potential give
no bound state for $V_\mathrm{I}$,
a shallow bound state $E_B \simeq 0.63~\mathrm{MeV}$ for
$V_\mathrm{II}$,\footnote{In Ref.~\cite{Morita:2016auo}
there is a typo in the binding energy with $V_\mathrm{II}$+Coulomb potential.
The value of $6.3~\mathrm{MeV}$ shown in Table I of Ref.~\cite{Morita:2016auo}
should be corrected to $0.63~\mathrm{MeV}$.},
and a deep bound state $E_B \simeq 26.9~\mathrm{MeV}$ for $V_\mathrm{III}$.

We find that the potential with nearly physical quark masses
is between $V_\mathrm{II}$ and $V_\mathrm{III}$;
the attraction becomes stronger with smaller quark masses,
but not as attractive as $V_\mathrm{III}$.
Consequently, the $p\Omega$ correlation function shown in this work
is also between those with $V_\mathrm{II}$ and $V_\mathrm{III}$
shown in~\cite{Morita:2016auo}.

%
\end{document}